\documentclass[11pt,a4paper,twoside]{article}
\newif\ifdraft \global\drafttrue
\def\production{\global\draftfalse}
\production

\usepackage[T1]{fontenc}
\usepackage[latin1]{inputenc}
\usepackage{a4wide}
\usepackage{times}
\usepackage{graphicx}
\usepackage{color}
\usepackage{fancyhdr}
\usepackage{latexsym}
\usepackage{amsmath}
\usepackage{amsthm}
\usepackage{bm}
\usepackage{bbm}
\usepackage{amssymb}
\usepackage{amsfonts}
\usepackage{enumitem}
\usepackage{cancel}
\usepackage{array}
\usepackage[colorlinks = true, linkcolor = teal, urlcolor = teal, citecolor = violet]{hyperref}

\usepackage{tikz}
\usetikzlibrary{positioning}
\usetikzlibrary{math}
\usetikzlibrary{patterns}

\usepackage{cleveref}
\usepackage{mathtools}
\usepackage{authblk}

\ifdraft
	\usepackage{showidx}
		\usepackage{showkeys}
\fi

\usepackage{xargs}

\newcounter{smallarabics}

\newcounter{smallroman}

\newcommand{\ben}{\begin{enumerate}[{\rm (1)}]}
\newcommand{\een}{\end{enumerate}}

\newtheorem{theorem}{Theorem}[section]
\newtheorem{proposition}[theorem]{Proposition}
\newtheorem{lemma}[theorem]{Lemma}

\newtheorem{definition}[theorem]{Definition}
\newtheorem{remark}[theorem]{Remark}
\newtheorem{remarks}[theorem]{Remarks}

\def\rr{{\mathbb R}}
\def\zz{{\mathbb Z}}
\def\cc{{\mathbb C}}
\def\nn{{\mathbb N}}

\def\clos{\mathrm{cl\,}}
\renewcommand{\sp}{\mathrm{sp}\,}

\def\proof{\noindent \textbf{Proof.}\ \ }

\def\cH{{\cal H}}
\def\ch{\mathfrak{h}}

\def\cR{{\cal R}}

\def\cB{{\cal B}}

\def\l{{\rm l}}

\def\b{{\rm b}}

\def\i{{\rm i}}
\def\alphab{{\bm \alpha}}
\def\betab{{\bm \beta}}
\def\zetab{{\bm \zeta}}
\newcommand{\eb}{{\bm e}}

\def\phib{\bm \phi}
\def\varphib{\bm \varphi}
\def\rhot{{\tilde\rho}}
\def\pp{{\mathbb P}}
\def\ee{\mathbb E}

\def\id{{\rm id}}

\newcommand{\e}{\mathrm{e}}
\renewcommand{\i}{\mathrm{i}}
\renewcommand{\d}{\mathrm{d}}
\renewcommand{\Re}{\mathop\mathrm{Re}}
\renewcommand{\Im}{\mathop\mathrm{Im}}
\newcommand{\strip}{\mathfrak I}
\newcommand{\hull}{\mathfrak C}

\newcommand{\norme}[1]{\|#1\|}
\newcommand{\module}[1]{|#1|}

\def\one{{\mathbbm 1}}
\def\ind{\mathbbm{1}}
\def\Dom{{\rm Dom}\,}
\def\eq{{\rm eq}}
\def\cS{{\cal S}}

\newcommand{\Blt}{\hyperlink{Blt}{$\bm{\mathrm{B(}}{\alpha_0,\theta_0}\bm{\mathrm{)}}$}}
\newcommand{\Blbt}{\hyperlink{Blbt}{$\bm{\mathrm{B_\betab(}}{\alpha_0,\theta_0}\bm{\mathrm{)}}$}}

\newcommand{\SFTL}{\hyperlink{SFTL}{\textbf{\textup{SFTL}}}}
\newcommand{\SFUV}{\hyperlink{SFUV}{$\bm{\mathrm{SFUV(}\gamma_0\mathrm{)}}$}}
\newcommand{\SFtwo}{\hyperlink{SFtwo}{\textbf{\textup{SF1}}}} 
\newcommand{\SFthree}{\hyperlink{SFthree}{\textbf{\textup{SF2}}}} 
\newcommand{\EBBTL}{\hyperlink{EBBTL}{\textbf{\textup{EBBTL}}}}
\newcommand{\TL}{\hyperlink{TL}{$\bm{\mathrm{TL}}$}}
\newcommand{\LT}{\hyperlink{LT}{$\bm{\mathrm{LT(}}\alpha_0\bm{\mathrm{)}}$}}
\newcommand{\LR}{\hyperlink{LR}{$\bm{\mathrm{LR(}}\alpha_0,\delta\bm{\mathrm{)}}$}}

\def\iL{^{(L)}}

\def\tr{{\rm tr}}

\newcommand{\sys}{\mathcal S}

\def\dGa{\d\hspace{-0.05em}\Gamma}

\usepackage{xparse}
\ExplSyntaxOn
\NewDocumentCommand{\mref}{m}{\quinn_mref:n {#1}}
\seq_new:N \l_quinn_mref_seq
\cs_new:Npn \quinn_mref:n #1
 {
 \seq_set_split:Nnn \l_quinn_mref_seq { , } { #1 }
 \seq_pop_right:NN \l_quinn_mref_seq \l_tmpa_tl
 (  \seq_map_inline:Nn \l_quinn_mref_seq
  { \ref{##1},\nobreakspace }  \exp_args:NV \ref \l_tmpa_tl  ) }
\ExplSyntaxOff

\begin{document}

\title{Heat conservation and fluctuations between quantum reservoirs in the Two-Time Measurement picture}
\author[1]{T. Benoist}
\author[2]{A. Panati}
\author[3]{Y. Pautrat}

\affil[1]{ Institut de Math\'ematiques de Toulouse, UMR5219, Universit\'e de Toulouse, CNRS, UPS IMT, F-31062 Toulouse Cedex 9, France}
\affil[2]{Aix Marseille Univ, Universit\'{e} de Toulon, CNRS, CPT, Marseille, France}
\affil[3]{Laboratoire de Math\'ematiques d'Orsay\\
Univ. Paris-Sud, CNRS, Universit\'e
Paris-Saclay\\ 91405 Orsay, France}

\maketitle

{\small
\textbf{Abstract.} 
This work concerns the statistics of the Two-Time Measurement definition of heat variation in each reservoir of a thermodynamic quantum system. We study the cumulant generating function of the heat flows in the thermodynamic and large-time limits. It is well-known that, if the system is time-reversal invariant, this cumulant generating function satisfies the celebrated Evans--Searles symmetry. We show in addition that, under appropriate ultraviolet regularity assumptions on the local interaction between the reservoirs, it satisfies a translation-invariance property, as proposed in [Andrieux \emph{et al.} New J. Phys. 2009]. We particularly fix some proofs of the latter article where the ultraviolet condition was not mentioned. We detail how these two symmetries lead respectively to fluctuation relations and a statistical refinement of heat conservation for isolated thermodynamic quantum systems. As in [Andrieux \emph{et al.} New J. Phys. 2009], we recover the Fluctuation-Dissipation Theorem in the linear response theory, short of Green--Kubo relations. We illustrate the general theory on a number of canonical models.}
\thispagestyle{empty}

\tableofcontents
\section{Introduction} 
In this article, we are interested in the heat exchange between different reservoirs forming an isolated quantum system. In the quantum setup, there exist multiple choices for the definition of the variation of heat in each reservoir, and different choices may have different mathematical properties and physical relevance. We consider the Two-Time Measurement (TTM) definition (which was introduced initially in \cite{Kur,Tas} for other thermodynamic quantities, see the reviews \cite{JOPP,EHM,CHT}). Within this definition we establish a statistical formulation of the heat conservation. We will see that this is a more subtle problem for quantum systems than for their classical counterparts. 

In this introduction we present our results informally in the case of a system consisting of two reservoirs and start by ignoring the classical/quantum distinction. Assume that the heat dumped in reservoir $i$ between times $0$ and $t$ is represented by a quantity $Q_j$. We describe its expected properties. Since the system is assumed isolated, one expects the first law of thermodynamics to hold in the form $Q_1+Q_2=W$ with $W$ the work put in by turning on the interaction between the reservoirs. If the work can be neglected, we then get an expression of the conservation of heat: $Q_1+Q_2$ is negligible with respect to the elapsed time in the experiment. One also expects a second law to hold in the form $\beta_1Q_1+\beta_2Q_2\geq 0$, where $\beta_1$ and $\beta_2$ are the inverse temperatures, stating that the entropy production is non negative. When $Q_1$, $Q_2$ are average thermodynamic quantities these two laws are easily proved from microscopic classical or quantum models.

Looking at finer properties of thermodynamic quantities leads to a description of $Q_1$, $Q_2$ and $W$ as random variables. First, one expects the thermodynamic system to be ergodic; that is, for a long enough experiment, one expects the time average of heat exchanges $Q_j$ to be equal to their mean value. In probabilistic terms, $Q_1$ and $Q_2$ should obey a law of large numbers:
\[\lim_{t\to \infty} \frac1t\,Q_1=\langle \Phi_1\rangle_+\quad \lim_{t\to \infty} \frac1t\,Q_2=\langle \Phi_2\rangle_+\]
where $\langle \Phi_j\rangle_+$ is the limit of the average heat flux $\frac1t\langle Q_j\rangle_t$ for $i=1,2$. As discussed above, if work can be neglected, one expects the relations $\langle \Phi_1\rangle_++\langle \Phi_2\rangle_+=0$ and $\beta_1\langle\Phi_1\rangle_++\beta_2\langle\Phi_2\rangle_+\geq 0$, from which we deduce that in the long time limit, heat flows from the hot to the cold reservoir. Looking at the first order of fluctuations around the law of large numbers, one expects a central limit theorem property, i.e.\ that the joint distribution of $(Q_1-t\langle \Phi_1\rangle_+)/\sqrt{t}$ and $(Q_2-t\langle \Phi_2\rangle_+)/\sqrt{t}$ converges to that of centered Gaussian random variables $\xi_1$ and $\xi_2$ such that $\xi_1=-\xi_2$ with probability $1$ or, equivalently, $(1,1)$ is in the kernel of their covariance matrix. A description of fluctuations beyond the central limit theorem typically uses a large deviation principle, i.e.\ gives a so-called rate function $I(q_1,q_2)$ such that the probability that $(\frac1tQ_1,\frac1tQ_2)$ is close to some value $(s_1,s_2)$ decays exponentially in time with rate $I(s_1,s_2)$. The route to a large deviation principle is usually through the G\"artner--Ellis theorem (see \cite{Ellis,DZ}). The latter is based on the study of the limit cumulant generating function $\chi_+(\alpha_1,\alpha_2)=\lim_{t\to\infty}\frac1t\log\chi_t(\alpha_1,\alpha_2)$, where 
\[\chi_t(\alpha_1,\alpha_2)=\mathbb E_t(e^{-\alpha_1 Q_1-\alpha_2Q_2})=\int_{\rr^2}e^{-\alpha_1q_1-\alpha_2q_2}\, \d\pp(q_1,q_2)\]
is the moment generating function of $(Q_1,Q_2)$ or Laplace transform of $\mathbb P_t$. Remark that the distribution $\pp_t$ is completely characterized by the function $\chi_t$ (see \cite{Bil}). It is now common knowledge that properties of $\chi_t$ can encode thermodynamic principles: assuming that the interacting dynamics is time-reversal invariant, $\pp_t$ should satisfy a transient fluctuation relation (see \cite{RM} or \cite{JOPP} and references therein for a detailed presentation of fluctuation relations) in the form:
\[\d\pp_t(q_1,q_2)=\e^{-(\beta_1q_1+\beta_2q_2)}\, \d\pp_t(-q_1,-q_2),\]
and an equivalent formulation is that the moment generating function verifies the Evans--Searles symmetry (see \cite{ES})
\begin{equation} \label{eq_transientfluctuationintro}
 \chi_t(\alpha_1,\alpha_2)=\chi_t(\beta_1-\alpha_1,\beta_2-\alpha_2).
\end{equation}
Given this symmetry, Jensen's inequality implies directly the second law of thermodynamics. Hence, the Evans--Searles symmetry \eqref{eq_transientfluctuationintro} can be seen as a statistical refinement of the second law. Let us emphasize that this symmetry holds for finite time $t$ (it is therefore called the transient fluctuation relation) and that this statistical refinement of the second law therefore does not require the introduction of large deviation theory.

It turns out that the properties related to the conservation of heat are not so trivial to express. Let us describe our approach by making the strongest sensible assumption on the maximal total heat input $Q_1+Q_2$, 
that it is bounded almost surely or namely that there exists $C>0$ such that $\pp_t$ almost surely one has $|Q_1+Q_2|\leq C$, for any time $t$. Then we obtain an almost translation symmetry in the moment generating function parameters: for any time and any $\theta\in \rr$,
\begin{equation} \label{eq_encadrementcasborneintro}
 \e^{-|\theta| C}\chi_t(\alpha_1,\alpha_2)\leq \chi_t(\alpha_1+\theta,\alpha_2+\theta)\leq e^{|\theta| C}\chi_t(\alpha_1,\alpha_2).
\end{equation}
Remark that this is not a translation symmetry since we do not have equalities. However, since $C$ does not depend on time, it turns into a symmetry for the asymptotic cumulant generating function $\chi_+=\lim_{t\to\infty}\frac1t\log\chi_t$.

If $\chi_+(\alpha_1,\alpha_2)$ is defined then trivially \eqref{eq_transientfluctuationintro} implies $\chi_+(\alpha_1,\alpha_2)=\chi_+(\beta_1-\alpha_1,\beta_2-\alpha_2)$, and \eqref{eq_encadrementcasborneintro} implies the new relation $\chi_+(\alpha_1+\theta,\alpha_2+\theta)= \chi_+(\alpha_1,\alpha_2)$ for all $\theta$. If $\chi_+$ is defined on $\rr^2$ then this translates into properties of the large deviations rate function $I$: the first relation satisfied by $\chi_+$ implies $I(s_1,s_2)=I(-s_1,-s_2)-(\beta_1s_1+\beta_2s_2)$, and the second implies $I(s_1,s_2)=\infty$ if $s_1+s_2\neq 0$, meaning that the probability of an event where conservation does not hold, such as $|(Q_1+Q_2)/t|>\epsilon$, decays superexponentially with time. In the present case this last statement would be a trivial consequence of the almost sure bound $|Q_1+Q_2|\leq C$ (once it is assumed that a large deviation principle holds for the pair $(Q_1,Q_2)$) without need for a detour by the cumulant generating function $\chi_+$.

However, the almost sure bound and the full translation symmetry may not always hold, i.e.\ even when $\chi_+(\alpha_1,\alpha_2)$ is defined, one may not have $\chi_+(\alpha_1,\alpha_2)=\chi_+(\alpha_1+\theta,\alpha_2+\theta)$. The extent of validity of this relation will depend on the total heat introduced by turning on the interaction. It is trivially verified by classical systems with bounded interaction between the reservoirs (see \cite[Proposition 2.3]{BPR}). Already for classical models of harmonic oscillators, it holds only for $\theta$ in a bounded interval: in that case the interaction is not bounded and there exists $\theta_0>0$, such that for any $\alpha_1,\alpha_2$ small enough, for any $\theta\in[-\theta_0,\theta_0]$, $\chi_+(\alpha_1,\alpha_2)=\chi_+(\alpha_1+\theta,\alpha_2+\theta)$, but one cannot take $\theta_0$ to infinity (see \cite{benoist2017energy,damak2019detailed}). Thankfully, this bounded translation symmetry still implies some bound on the rate function: the probability that heat is not conserved does not decay superexponentially, but exponentially with rate proportional to $\theta_0$. Namely, if there exists $\theta_0>0$ such that for any $\alpha_1, \alpha_2$ small enough and any $\theta\in [-\theta_0,\theta_0]$ one has $\chi_+(\alpha_1,\alpha_2)=\chi_+(\alpha_1+\theta,\alpha_1+\theta)$, then $I(s_1,s_2)\geq \theta_0\,|s_1+s_2|$, and in particular the probability of $|(Q_1+Q_2)/t|>\epsilon$ decays exponentially with time.

Summarizing, when concerned with the laws of thermodynamics beyond the law of large number or the central limit theorem, it is useful to study the asymptotic cumulant generating function $\chi_+$ of the heat fluxes. We expect two symmetries in its parameters: the Evans--Searles symmetry is related to the second law through the fluctuation relation, and the translation symmetry induces a control of heat conservation violations. Interestingly, using jointly both of these symmetries, one can prove the fluctuation--dissipation theorem short of Green--Kubo relations (see \cite{AGMT} and \Cref{subsec_linearresponse}). As already mentioned, for classical system with bounded inter-reservoirs interaction, proving both symmetries is easy; for unbounded interactions the approach has to be model-dependent but no conceptual difficulty arises.  

For quantum systems, however, even with bounded interactions the very definition of the random variables $Q_1$ and $Q_2$ raises difficulties and different definitions exist in the literature. Let us introduce some notation: let $H_1$, $H_2$ be the free Hamiltonians of the two reservoirs and let $V$ denote the bounded interaction between them. The Hamiltonian governing the dynamics is $H=H_0+V$ with $H_0=H_1+H_2$ and $(\tau^t)_t$ is the associated Heisenberg dynamics, i.e.\ the evolved version $A_t$ of the observable $A$ is $A_t=\tau^t(A):=\e^{\i t H}A\e^{-\i t H}$. Let $\rho$ denote the initial state of the system; namely $\rho$ is a density matrix such that the expectation value of an observable $A$ is $\tr(\rho A)$ (which we denote $\rho(A)$ for concision). A ``naive quantization'' approach would define for $i=1,2$ the variable $Q_j$ as proceeding from a measurement of the observable $\Delta_t H_j=H_{j,t}-H_j$, and similarly the work variable $W$ as proceeding from a measurement of $\Delta_t V=V-\tau^t(V)$. One then benefits from the obvious identity $\Delta_t H_1+\Delta_t H_2=\Delta_t V$, but the fluctuation relation \eqref{eq_transientfluctuationintro} fails (see \cite[Exercices 3.3 \& 6.1]{JOPP}) and the thermodynamic quantities are hard to interpret physically. The TTM approach defines thermodynamic quantities through the following \emph{Gedankenexperiment}: if one measures $H_1$ and $H_2$ at time $0$ with outcomes $h_1$ and $h_2$ (remark that $H_1$ and $H_2$ commute and can therefore be simultaneously measured), lets the (post-measurement) system evolve for a time $t$, and measures $H_1$ and $H_2$ again with outcomes $h_1'$ and $h_2'$, then $Q_1$ and $Q_2$ are defined as the (random) differences $h_1'-h_1$ and $h_2'-h_2$. This definition is easy to interpret physically and ensures the validity of the fluctuation relation (see below), but the equality $Q_1+Q_2=W$ does not hold beyond the first two moments (see \cite[\S3.5]{JOPP}), so that a uniform bound for $W$ does not tell us anything about $Q_1+Q_2$. Similarly, the inequality
\begin{equation} \label{eq_bornetriviale}
\big| \rho\big(\tau^t(H_1)-H_1\big)+ \rho\big(\tau^t(H_2)-H_2\big)\big|\leq 2 \|V\|
\end{equation}
leads to $\ee(Q_1)+\ee(Q_2)\leq 2\norme V$ but this does not prove the almost-sure boundedness of $Q_1+Q_2$. Thankfully, as proved in \cite{BJPPP,BPR}, under some ultraviolet regularity conditions on the interaction one can control the exponential decay of the probability of violation of heat conservation; these ultraviolet regularity conditions are actually shown in \cite{BPR} to be essentially necessary to control the tails of the total heat distribution and some examples with bounded interactions are provided where the fourth moment $\mathbb E_t((Q_1+Q_2)^4)$ of the total heat is infinite for almost every time $t$.

In the present article we work with the TTM definition of thermodynamic quantities and study the statistics of the pair $(Q_1,Q_2)$. We introduce ultraviolet regularity conditions generalizing those of \cite{BJPPP,BPR}, that allow us to prove a bounded translation symmetry of the cumulant generating function. Our proof fixes an error in the derivation of the translation symmetry provided in \cite{AGMT}, where the authors overlooked the necessity of ultraviolet regularity conditions (see \Cref{rk:error_AGMT}). We then present and derive rigorously the previously mentioned consequences of the symmetries satisfied by the cumulant generating function (see \Cref{sec_OpenQuantumSystems}), and derive the fluctuation--dissipation theorem short of Green--Kubo formula (\Cref{theo_onsager}) At the end of the article we give some examples of application to standard models of quantum statistical mechanics (see \Cref{sec-models}). Before we enter into a more rigorous discussion, we conclude this section with an informal presentation of our assumptions and results. 

Our assumption is essentially that for some strictly positive $\alpha_0$ both of the following bounds hold:
\begin{gather}\label{eq_regularity}
\sup_{\alpha_1,\alpha_2\in[-\alpha_0,+\alpha_0]}\big\|\e^{+\frac12(\alpha_1 H_1 +\alpha_2 H_2)} \,V \,\e^{-\frac12(\alpha_1 H_1 +\alpha_2 H_2)}\big\|<\infty\\
\sup_{\alpha_1,\alpha_2\in[-\alpha_0,+\alpha_0]}\big\|\e^{+\frac12((\alpha_1+\beta_1) H_1 +(\alpha_2+\beta_2) H_2)} \,V \,\e^{-\frac12((\alpha_1+\beta_1) H_1 +(\alpha_2+\beta_2) H_2)}\big\|<\infty.
\end{gather}
 We call this assumption an ultraviolet regularity as in paradigmatic models of non-equilibrium statistical mechanics the condition translates directly to an ultraviolet regularity condition (see \Cref{subsec_spinfermion} and \cite{BPR}). It can also be understood through a golden rule approximation of the energy transitions induced by the interaction: in such approximation the transition rate between two $H_0$ eigenstates $|\Psi_E\rangle$ and $|\Psi_{E'}\rangle$ of respective energy $E$ and $E'$ is essentially given by $T(E,E')=|\langle \Psi_{E'}|V\Psi_{E}\rangle|^2$. Then the uniform bounds of equation \eqref{eq_regularity} imply for example that $T(E,E')=o(e^{-\alpha_0 |E-E'|})$, namely, transitions towards high energy (i.e.\ ultraviolet) states are exponentially suppressed.
Although obviously a stronger condition than the boundedness of $V$ which was assumed to derive \eqref{eq_bornetriviale}, this ultraviolet regularity condition holds in many models of physical interest; see \Cref{sec-models}. As already mentioned, the results of \cite{BPR} show that this ultraviolet regularity condition is essentially necessary for the probability of violation of heat conservation to be exponentially suppressed. The emerging underlying picture is that, in the quantum setting, total heat fluctuations are sensitive to energy transitions induced by the interaction (i.e.\ ultraviolet regularity of~$V$), rather than to the interaction strength (i.e.\ $\|V\|$).

In the present article, we only deal with bounded interactions. Our results thus do not apply to reservoirs of bosons. Nevertheless the intuition behind our proofs should apply to models involving unbounded interaction. Though, in this case, as for classical models with unbounded interaction, the strength of the interaction may also limit the size of the interval on which the translation symmetry is valid. A first approach to the questions we deal with here but for unbounded $V$ can be found in \cite[\S4.2 and \S4.3]{BPR}.

We choose to describe infinitely extended systems through their approximations by finite dimensional systems. Infinitely extended reservoirs are required to observe non-trivial thermodynamic behavior, but their description relies typically on infinite-dimensional operator algebras, and the related theory of e.g.\ states and dynamics is technically demanding. We can however state our assumptions exclusively in terms of generating functions associated with finite-dimensional approximations. Although such a route may have its limitations (see \cite{JOPP} for an in-depth discussion), it allows us to bypass heavy algebraic machinery and make it accessible to a larger audience. Our mathematical analysis relies essentially on simple trace inequalities that allow us to give relevant bounds. The thermodynamic limit is only performed at the level of generating functions, and therefore of probability distributions. The probabilistic tools involved are elementary, except possibly for the G\"artner-Ellis theorem, used to obtain a large deviation principle. Note that \cite{BPR} takes a different approach and studies directly infinitely extended systems via algebraic and analytic tools. 

\smallskip
The article is organized as follows. In \Cref{sec_confined} we introduce our main objects of interest in a finite-dimensional setting, i.e.\ for confined systems, and derive various bounds for generating functions from simple trace inequalities. In \Cref{sec_OpenQuantumSystems} we specialize to thermodynamic quantum systems composed of several thermal reservoirs, and 
establish results concerning the joint distribution for the heat variations in the different reservoirs: we show a law of large numbers, a central limit theorem and a large deviation principle. We also recover the Fluctuation-Dissipation Theorem (short of Green--Kubo relations), and discuss the role of ``small'' systems, i.e.\ systems whose Hilbert space stays finite dimensional in the thermodynamic limit. In \Cref{sec-models}, we present various models for which we discuss our ultraviolet regularity assumptions and those of our results that apply.

\bigskip\noindent
{\bf Acknowledgments.} 
The research of T.B.\ was supported by ANR project RMTQIT (Grant No. ANR-12-IS01-0001-01), LabEx CIMI (ANR-11-LABX-0040-CIMI within the program ANR-11-IDEX-0002-02 ) and by ANR contract StoQ ANR-14-CE25-0003-0. The research of A.P.\ was partially supported by ANR project SQFT (ANR-12-JS01-0008-01) and ANR grant NonStops (ANR-17-CE40-0006-01, ANR17-CE40-0006-02, ANR-17-CE40-0006-03), and part of her work was done during a CNRS leave at CRM, Montreal, UMI 3457. The research of Y.P.\ was partly supported by ANR contract StoQ ANR-14-CE25-0003-0 and ANR grant NonStops (ANR-17-CE40-0006-01, ANR17-CE40-0006-02, ANR-17-CE40-0006-03), and part of his work was done during a CNRS leave at CRM, Montreal, UMI 3457. All three authors wish to thank the mathematics and statistics department of McGill University, where this work was initiated, and Vojkan Jak\v{s}i\'c for suggesting the problem.

\section{Definitions for confined systems and basic bounds} \label{sec_confined}
In this section, we consider systems with a finite number of degrees of freedom. This allows us to define without analytical difficulties the random variables corresponding to thermodynamic quantities in the Two-Time Measurement picture and to derive relevant inequalities for their joint generating function from basic trace and norm inequalities for matrices.

\subsection{Setup: observables and measurement} \label{subsec_ConfinedSetup}
We first recall the basic formalism of quantum mechanics: a physical system is described by a triple $(\cH, H, \rho)$ where $\cH$ is an Hilbert space, $H$ a self-adjoint operator on $\cH$ and $\rho$ is a density matrix, i.e.\ a nonnegative operator $\rho$ on $\cH$ with trace $1$ (we will use the same symbol $\rho$ for the density matrix and the linear form $A\mapsto \tr(\rho A)$). We consider a \emph{confined system} in the sense that $\dim \cH<\infty$. In this setting, denoting by $\cB(\cH)$ the algebra of (bounded) linear operators on $\cH$, the physical observables can be identified with self-adjoint elements $A$ of $\cB(\cH)$. The dynamics of the system is dictated by the operator $H$ which is interpreted as energy observable and is called the \emph{Hamiltonian}. We will work in the Heisenberg picture where an observable evolved to time $t$ is given by $A_t=\e^{+\i t H}A\,\e^{-\i t H}$ while the state $\rho$ stays constant.

We also recall the standard description of measurement: the measurement of an observable $A$ has random outcomes taking values in the spectrum $\sp A$ with probability depending on the pre-measurement state $\rho$. More precisely, if we write $A=\sum_{a\in \sp A} a P_a$ the spectral resolution of $A$, then a measurement of the observable $A$ on a system with pre-measurement state $\rho$ will return the value $a\in \sp A$ with probability $\tr(P_a \rho)$. In particular, the expectation and variance of the measurement outcome are respectively $\rho(A)$ and~$\rho\big((A-\rho(A))^2\big)$. Conditioned on outcome $a$, the post-measurement state is $P_a\rho P_a/\tr(P_a\rho)$.

We are interested in systems consisting of several ``reservoirs'' $\cR_1,\ldots,\cR_\ell$. Once again we will approximate physical reservoirs by confined systems; this will be formalized in the next definition.
 Therefore, we view our confined system as consisting of different subsystems. Typically, the Hilbert space can be written as product of Hilbert spaces $\cH=\cH_1 \otimes \ldots \otimes \cH_\ell$, and to each reservoir one can associate a "free" energy $H_j$, $j=1,\ldots, \ell$, generating the dynamics of subsystem $\cR_j$. For the full system we consider two different Hamiltonians: one is $H_0=\sum_{j=1}^\ell H_j$ which we call the \emph{free Hamiltonian}, another is  $H=H_0+V$ which we call the \emph{full Hamiltonian}. The observable $V$ therefore represents the interaction between the different $\cR_j$.

For compactness, we set  $\bm E=(H_1,\ldots, H_\ell)$ and we give the following definition.

\begin{definition}
A \emph{confined multisystem} is a quadruple $(\cH, \rho, {\bm E}, V)$, with $\cH$ a finite-dimensional Hilbert space, ${\bm E}$ a finite set of commuting observables, $\rho$ a density matrix and $V$ an additional observable. 
\end{definition}



In what follows,  the nature of the reservoir  $\cR_j$ is irrelevant and the decomposition of  the  Hilbert space only appears in the fact that we consider a set  of commuting observables $\bm E=(H_1,\ldots, H_\ell)$. Therefore, the analysis in the remainder of the paper can be extended to any system formally satisfying the above definition. Note that, with the above definition, we do not need to specify the nature of the interaction, and in particular whether the reservoirs interacts directly or through another system. This will be relevant in the thermodynamic limit, see \Cref{remark_smallsys} and \Cref{subsec_smallsystem}.

\textbf{Notation:} 
We follow the convention that bold letters (e.g.\ $\bm{A,a,\phi}$) are used for $\ell$-tuples. If $\bm a, \bm b \in \cc^\ell$, $\bm a .\bm b=\sum_{j=1}^\ell a_j b_j$ is the bilinear quadratic form extending to $\cc^\ell$ the canonical scalar product of $\rr^\ell$. From now on, we always write $\sum_j$ for $\sum_{j=1}^\ell$, and use notation such as $\alphab.\bm{E}=\sum_j \alpha_j H_j$ for $\alphab \in\cc^\ell$. In addition, we denote $|\alphab|=(|\alpha_1|,\ldots,|\alpha_\ell|)$ and $\norme{\alphab}=\sup_{j=1,\ldots\ell}|\alpha_j|$.

A particular family of states will be relevant in confined multisystems. They are of the form
\begin{equation} \label{eq_GibbsState}
\rho_{\bm{\beta}}=Z^{-1}\,\e^{-\sum_j \beta_j H_j}=Z^{-1}\,\e^{-\bm\beta . \bm E}\qquad \mbox{with } Z=\tr(\e^{-\sum_j \beta_j H_j})=\tr(\e^{-\bm\beta . \bm E}),
\end{equation}
where $\bm{\beta}=(\beta_1,\ldots,\beta_\ell)$ and $\beta_j>0$ for all $j$. A state of this form is called a \emph{multi-thermal state} with respect to inverse temperatures $\betab$. If $\rho$ is a multi-thermal state, then initially each subsystem is in thermal equilibrium, and if two of the inverse temperatures $\beta_j$ are different, then the temperature differential will result, in the thermodynamic and large-time limit, in the onset of steady heat fluxes across the system. Our goal is to study the statistical properties of those fluxes.

Taking a ``naive'' quantization picture, these fluxes are described by the observables $\Phi_j=\i [H,H_j]$. One has immediately
\[H_{j,t}-H_j=\int_0^t\Phi_{j,s}\,\d s\]
so the convention is that heat flowing into a subsystem $\cR_j$ corresponds to positive values of $\Phi_j$. 
\subsection{Setup: Two-Time Measurement picture} \label{subsec_FCS}

In this section we fix a confined multisystem $(\cH, \rho, \bm E, V)$ and construct the joint probability distribution $\pp_t$ of heat variation rates into the different reservoirs, according to the Two-Time Measurement picture.

The set up is the following. The commuting observables $H_1, H_2, \ldots, H_\ell$ are measured simultaneously a first time. For short we say we measure $\bm E$. The system evolves for a time $t$, then  $\bm E$ is measured again.
We denote by $\sp \bm E$  the possible outcomes of a measurement i.e.
\[\sp \bm E = \{\bm e = (e_1,\ldots, e_\ell),\, e_j\in\sp H_j \mbox{ for }j=1,\ldots,\ell\}.\]
Let $(P_{\bm e})_{\bm e \in \sp \bm E}$ be the commuting family of projectors such that for any $j=1,\ldots, \ell$, one has $H_j=\sum_{\bm e \in \sp \bm E }e_j \,P_{\bm e}$. If the system is initially in the state~$\rho$, then the outcome of a measurement of $\bm E$ will be $\bm e \in \sp \bm E$ with probability $\tr(P_{\bm e}\rho P_{\bm e})$, and after the measurement the system is in the state $\tfrac{P_{\bm e}\rho P_{\bm e}}{\tr(P_{\bm e}\rho P_{\bm e})}$.
We can equivalently write the above probability and post-measurement state as $\tr(\rhot\, P_{\bm e})$ and $\rhot\, P_{\bm e}/\tr(\rhot \,P_{\bm e})$, where $\rhot$ is the \emph{a priori state} with respect to $\bm E$ defined by
\begin{equation} \label{eq_defrhotilde}
\rhot =\sum_{\bm e\in\sp{\bm E}} P_{\bm e} \rho P_{\bm e}.
\end{equation}
In many situations of interest, and in particular whenever $\rho$ is a multi-thermal state, $\rho$ will commute with $\bm E$ and we will therefore have $\rho=\rhot$.
Assume now that, after measuring $\bm E$ at time $0$, with outcome $\bm e$, we let the system evolve for a time $t$ before making a second measurement of $\bm E$. The post-measurement state evolves, after a time $t$, into $\tfrac{\e^{-\i t H}\rhot\, P_{\bm e}\, \e^{+\i t H}}{\tr(\rhot \,P_{\bm e})}$. The second measurement of $\bm E$ then gives $\bm e^\prime$ with probability $\frac{\tr (\e^{-\i t H}\rhot \, P_{\bm e}\,\e^{+\i t H}P_{\bm e^\prime})}{\tr (\rhot \, P_{\bm e})}$. The probability that these two measurements give $\bm e$ then $\bm e^\prime$ is therefore $\tr (\e^{-\i t H}\rhot \, P_{\bm e}\,\e^{+\i t H}P_{\bm e^\prime})$. The joint law of the heat fluxes towards the different reservoirs is the induced probability measure ${\pp}_{t}$ on $\rr^\ell$ of the vector ${\bm\phi}=({\bm e^\prime}-{\bm e})/t$ and describes the rate of change of ${\bm E}$ between the two measurements. More precisely, for ${\bm \varphi}\in\rr^\ell$,
\begin{equation} \label{eq_defFCS}
{\pp}_{t}(\varphib)=\sum_{\bm e, \bm e^\prime \in \sp \bm E} \ind_{\bm e^\prime - \bm e = t \varphib}\, \tr \big(\e^{-\i t H}\rhot \,P_{\bm e}\e^{+\i t H}P_{\bm e^\prime}\big).
\end{equation}
The measure ${\pp}_{t}$ is concentrated on the set $(\sp{\bm E}-\sp{\bm E})/t$. The random variable $\phib$ is the canonical coordinate mapping $\varphib\mapsto\varphib$. We denote by $\ee_t$ the expectation with respect to $\pp_t$. We define a map $\chi_{t}$ from $\cc^\ell$ to $\cc$:
\begin{equation} \label{eq_defcorr}
\chi_{t}(\alphab)= \ee_t\big(\exp (-t \alphab . \phib)\big)=\sum_{\varphib}\e^{-\sum_j t \alpha_j\varphi_j}\,{\pp}_{t}(\varphib).
\end{equation}

We will occasionally consider two additional random variables. The first is $\phi_0$, defined by 
\begin{equation*}
 \phi_0 :{\bm{\varphi}}\mapsto \bm 1.{\bm\varphi}=\sum_j \varphi_j.
\end{equation*}
The second is $\varsigma$, defined in the case where the state $\rho$ is multi-thermal with respect to inverse temperatures $\betab$ by
\begin{equation*}
 \varsigma :{\bm{\varphi}}\mapsto \bm \betab.{\bm\varphi}=\sum_j \beta_j\varphi_j.
\end{equation*}
It is clear that $\phi_0$ and $\varsigma$ are random variable emerging from the Two-Time Measurement of $H_0$ and (when the state $\rho$ is multi-thermal) $S=\sum_j \beta_j H_j+\log Z=-\log\rho$ respectively. Therefore, $\phi_0$ represents the rate of total heat absorption and $\varsigma$ the rate of change of total entropy in the Two-Time Measurement picture. Our statements below on the distribution of $\phib$ obviously translate into statements on the distributions of $\phi_0$ and $\varsigma$.

In the remainder of the section, we list some immediate properties of $\chi_{t}$ and establish a connection with the first two momenta of the heat fluxes.

First, one easily verifies from \eqref{eq_defFCS} and \eqref{eq_defcorr} that 
\begin{equation} \label{eq_defcorr2}
\chi_{t}({\bm \alpha})=\tr \big(\e^{-\i t H}\rhot\, \e^{+{\bm \alpha}. {\bm E}}\e^{+\i t H}\e^{-{\bm \alpha}. {\bm E}}\big)=\tr \big(\e^{-\frac12{\bm \alpha}. {\bm E}}\e^{-\i t H}\e^{+\frac12{\bm \alpha}. {\bm E}}\rhot\,\e^{+\frac12{\bm \alpha}. {\bm E}}\e^{+\i t H}\e^{-\frac12{\bm \alpha}. {\bm E}}\big).
\end{equation}
It is then clear that $\chi_{t}$ is an entire function of $\alphab$ and $\chi_{t}(\alphab)\in\rr_+$ for $\alphab\in\rr^\ell$. The restriction of $\alphab\mapsto\log \chi_{t}(\alphab)$ to $\rr^\ell$ is a convex function\footnote{We define a convex function as a map $f$ from $\rr^\ell$ to $]-\infty,+\infty]$ which is not $+\infty$ everywhere, and for $\alphab_1,\alphab_2\in\rr^\ell$ and $u\in[0,1]$ satisfies $f(u\alphab_1+(1-u)\alphab_2)\leq u f(\alphab_1) +(1-u) f(\alphab_2)$.}. The triangle and Cauchy-Schwarz inequalities imply respectively
\begin{gather}
|\chi_{t}(\alphab)|\leq \chi_{t}(\mathrm{Re}\,\alphab) \quad \mbox{for }\alpha\in\cc^\ell, \label{eq_triangle}\\
\chi_{t}(-\alphab)^{-1}\leq \chi_{t}(\alphab) \quad \mbox{for }\alpha\in\rr^\ell. \label{eq_CauchySchwarz}
\end{gather}

An easy computation shows that
\begin{equation} \label{eq_DrvChit0}
\begin{aligned}
\frac{\partial}{\partial{\alpha_j}} \chi_{t}(\alphab)_{|\alphab = \bm 0} &= - \tr\big(\rhot(H_{j,t}-H_j)\big),\\
\frac{\partial^2}{\partial{\alpha_j} \partial{\alpha_k}}\chi_{t}(\alphab)_{|\alphab = \bm 0} &= \tr\big(\rhot (H_{j,t}-H_j)(H_{k,t}-H_k)\big).
\end{aligned}
\end{equation}

Recalling that we denote by $\Phi_j=\i [H, H_j]$ the flux observable of $H_j$ in the ``naive'' quantization definition, we thus have
\begin{equation} \label{eq_DrvChit}
\begin{aligned}
\ee_t(\phi_j)&=-\frac{1}{t}\frac{\partial}{\partial{\alpha_j}}\chi_{{t}}(\alphab)_{|\alphab=\bm 0}=\frac{1}{t}\int_0^t\rhot(\Phi_{j,s})\,\d s, \\
\ee_t(\phi_j\phi_k)&=
\frac{1}{t^2}\frac{\partial^2}{\partial{\alpha_j} \partial{\alpha_k}}\chi_{t}(\alphab)_{|{\bm \alpha}=\bm 0}=\frac{1}{t^2}\int_0^t \!\! \int_0^t \rhot(\Phi_{j,r}\Phi_{k,s})\,\d r \,\d s, \\
\mathrm{cov}_{\pp_{t}}(\phi_j,\phi_k) &=\frac{1}{t^2}\int_0^t \!\! \int_0^t \rhot\big((\Phi_{j,r}-\rhot(\Phi_{j,r}))(\Phi_{k,s}-\rhot(\Phi_{k,s}))\big)\,\d r \,\d s.
\end{aligned}
\end{equation}
We will denote $\langle \Phi_j\rangle_t=\frac{1}{t}\int_0^t\rhot(\Phi_{j,s})\,\d s$. Assume temporarily that $\rho$ commutes with $\bm E$ (as is the case if e.g.\ $\rho$ is multi-thermal); then $\rho=\rhot$ and \Cref{eq_DrvChit0,eq_DrvChit} show that for all $j$ and $k$
\[\ee_t(\phi_j)= \frac{1}{t}\,\rho(H_{j,t}-H_j),\qquad \ee_t(\phi_j\phi_k)= \frac{1}{t^2}\,\rho\big((H_{j,t}-H_j)(H_{k,t}-H_k)\big),\]
and we recover the well-known fact (see \cite{DdRM}) that the heat fluxes defined in the Two-Time Measurement picture have the same first two moments as the distribution of the observable $(\bm E_t-\bm E)/t$. Using $H_t=H$, it implies in particular that
\begin{equation*}
\begin{aligned}
\sum_{j} \langle \Phi_j\rangle_t=\frac1t\, \rho(V-V_t)\quad\text{and}\quad
\sum_{j,k} \mathrm{cov}_{\pp_t}(\phi_j,\phi_k)=\frac{1}{t^2}\,\rho\big((V-V_t)^2\big).
\end{aligned}
\end{equation*}
It follows that
\begin{equation}\label{eq:bound_mean_cov}
\begin{aligned}
\Big|\sum_{j} \langle \Phi_j\rangle_t\Big|\leq \frac{2\|V\|}{t}\quad\text{and}\quad
\sum_{j,k} \mathrm{cov}_{\pp_t}(\phi_j,\phi_k)\leq\frac{4\|V\|^2}{t^2}.
\end{aligned}
\end{equation}

\begin{remark} One can also consider an additional set of commuting observables $(N_1, \ldots, N_\ell)$ that satisfy $[H_j, N_k]=0$ for all $j,k$ where $N_j$ is interpreted as the observable counting the number of particle in the $j$-th subsystem, and one can study the Two-Time Measurement of $(H_1, \ldots, H_\ell, N_1, \ldots, N_\ell)$ with respect to the state
\[
\rho=\frac{\e^{-\sum_j \beta_j (H_j-\mu_jN_j)}}{\tr (\e^{-\sum_j \beta_j (H_j-\mu_jN_j)})},
\]
where $\mu_j\in \rr$ is the chemical potential of the $j$-th subsystem. Our results extend directly to this setting and we will leave these extensions to the interested reader.
\end{remark}

In the rest of this section, we collect various bounds on the moment generating function $\chi_t$, which we will then use to study the thermodynamics of systems described as infinite-dimensional limits of confined multisystems.

\subsection{Transient fluctuation relation for confined systems} \label{subsec_EvansSearles}

As mentioned in the introduction, the Two-Time Measurement definition of thermodynamic quantities allows for an extension of the transient fluctuation relations to the quantum setting (\cite{Kur,Tas,TasMat}). Transient fluctuation relations are equivalent to a symmetry of the generating function, often called Evans--Searles symmetry (see \cite{ES}, and \cite{RM, JOPP} for more references). In the multi-reservoir context, these relations are often called \emph{exchange fluctuation relations} (\cite{Jar04, CHT, EHM}). For the reader's convenience, we present a precise statement about fluctuation relation and the corresponding symmetry in our context.

\begin{definition} \label{defi_TRI}
We say that the confined multisystem $(\cH,\rho, \bm E, V)$ satisfies time-reversal invariance if there exists an antilinear involution $C$ on $\cH$ that commutes with $H$, $\rho$, and all $H_j$'s. In other words, time-reversal invariance holds iff $\cH$ has an orthonormal basis in which the matrix elements of $H$, $\rho$, and all $H_j$'s are real.
\end{definition}

We then have:
\begin{proposition} \label{prop_EvansSearles} Let $(\cH,\rho,\bm E,V)$ be a time-reversal invariant confined multisystem, where $\rho$ is multi-thermal with respect to inverse temperatures $\betab$. Then 
\begin{equation}
\label{eq_ESsym_finite}
\chi_t({\bm \alpha})=\chi_t({\bm \beta-\alphab}) \ \mbox{ for any }\alphab\in\rr^\ell.
\end{equation}
or equivalently $\pp_t(\varphib)$ and $\pp_t(-\varphib)$ are mutually absolutely continuous and
\begin{equation} \label{eq_ESflucrel}
 \frac{{\pp}_t(+\varphib)}{{\pp}_t(-\varphib)}=\e^{+t\sum_j \beta_j \varphi_j}
\end{equation}
for any $\varphib\in\rr^\ell$ such that $\pp_t(\varphib)\neq 0$.\end{proposition}
Because of the choice of initial state, the proof is a trivial consequence of time-reversal invariance (see \cite[Proposition 3.9]{JOPP} for details).

\begin{remark}\label{remark_ESflucrel}
Relation \eqref{eq_ESflucrel} implies in particular $\sum_j \beta_j \langle \Phi_j\rangle_t\geq 0 $, which is the usual expression of the positivity of the (average) entropy production. To describe a more familiar form of fluctuation relation, denote by $e_t$ the moment generating function of~$\varsigma$, i.e.\ the function $\rr\ni \alpha\mapsto \ee_t(\e^{-\alpha \varsigma})$. Then relation \eqref{eq_ESsym_finite} implies 
 \[ e_t(\alpha)=e_t(1-\alpha),\]
which in turn is equivalent to $\pp_t(\varsigma=+s)$ and $\pp_t(\varsigma=-s)$ being mutually absolutely continuous and
\begin{equation*}
 \frac{\pp_t(\varsigma=+s)}{\pp_t(\varsigma=-s)}=\e^{+t s}
\end{equation*}
for any $s$ such that $\pp_t(\varsigma=s)\neq 0$.
 \end{remark}

\subsection{Bounds for confined systems} \label{subsec_bounds}

In this section we give relevant bounds on the generating function $\chi_t$ associated with the family $\bm E=(H_1,\ldots,H_\ell)$.  Note in particular that \Cref{cut} proves a relation corresponding to \Cref{eq_encadrementcasborneintro}. Detailed proofs are given in \Cref{sec_proofs_bounds}.
 
These bounds are immediate in the classical case when $V$ is bounded, but, as explained in the introduction, the situation is more subtle in the quantum TTM framework, as the fluctuations are controlled by ultraviolet properties of the interaction rather than its strength. In the confined case, however, this problem can not occur and classical proofs can be extended making use of trace inequalities (see \Cref{sec_proofs_bounds}). The choice of the constants in the bounds below becomes relevant only in the thermodynamic limit and is therefore discussed in \Cref{sec_OpenQuantumSystems} (see also \Cref{sec-models}).

Let $(\cH,\rho,\bm E,V)$ be a confined multisystem. 
For any $\alphab\in\rr^\ell$, let
\begin{equation} \label{eq_defValpha}
V_{\bm\alpha}= \e^{+\frac12\alphab.\bm E} V \e^{-\frac12\alphab.\bm E}.
\end{equation}
For $\alpha_0\in\,\rr_+$ we introduce the constant
\begin{equation}\label{Const-Rj}
 S(\alpha_0)=\sup_{\norme{\alphab}\leq \alpha_0} \|V_\alphab\|.
\end{equation}
Note that there cannot exist a finite uniform bound for $\sup_{\alphab\in \rr^\ell} \|\e^{+\frac12\alphab. \bm E}\,V\,\e^{-\frac12\alphab.\bm E}\|$ unless $V$ commutes with $\bm E$. We discard this situation as physically uninteresting since it would imply that all considered observables are constant (and, after the thermodynamic limit, that $V$ is non local). 

We start by giving estimates on $\chi_t(\alphab)$. The relevance of $S(\norme{\alphab})$ comes from the fact that $\alphab.\bm E$ and $H=H_0+V$ do not commute, so that the term $\e^{{+\frac12\bm \alpha}. {\bm E}}\e^{+\i t H}\e^{-\frac12{\bm \alpha}. {\bm E}}$ in the expressions \eqref{eq_defcorr2} for $\chi_t$ can essentially only be controlled by writing $\e^{+{\bm \alpha}. {\bm E}}\e^{+\i t H}\e^{-{\bm \alpha}. {\bm E}}= \e^{+\i t (H_0+V_\alphab)}$ with $V_\alphab$ non self-adjoint. This is made more precise in the following statement.
\begin{proposition} \label{prop_energycorr2}
For any $\bm\alpha$ in $\rr^\ell$ we have 
\[\e^{-2|t|S(\norme{\alphab})}\leq \chi_t(\alphab)\leq \e^{+2|t|S(\norme{\alphab})}. \]
\end{proposition}

Remark that, if we define 
\begin{equation} \label{eq_defSH0}
 S_{H_0}(\theta_0)=\sup_{|\theta|\leq \theta_0}\|\e^{+\frac12\theta H_0} \,V\,\e^{-\frac12\theta H_0}\|
\end{equation}
then a proof using similar arguments shows the time-independent bounds
\begin{equation}\label{eq_BJPPP_bound}
\e^{-2|\theta|{S_{H_0}(\theta)}}\leq\chi_t(\theta \bm 1)\leq \e^{+2|\theta|{S_{H_0}(\theta)}}.
\end{equation}
This was used in \cite{BJPPP} to derive results on the large deviations of the total heat flux random variable~$\varphi_0$ with respect to $\pp_t$. We state a similar result in \Cref{remark_BJPPP}.

Let $(\cH,\rho,\bm E,V)$ be a confined multisystem with $\rho$ multi-thermal with respect to inverse temperatures~$\betab$; in this subsection we prove a bound relating $ \chi_t({\alphab} + \theta{\bm 1})$ to $\chi_t({\alphab})$. 
We define for $\alpha_0\in\,\rr_+$
\[\cB(\alpha_0):=\{\alphab\in\rr^\ell : \alphab.\bm 1=0, \|\alphab\|<\alpha_0\},\]
and for $\theta_0\geq \alpha_0$,
\begin{equation} \label{Const-R}
S(\alpha_0,\theta_0)=\sup_{|\theta|\leq \theta_0}\ \sup_{\alphab\in\cB(\alpha_0)} \|V_{\alphab+\theta\bm 1}\| 
\end{equation}
and
\begin{equation} \label{Const-R2}
 S_\betab(\alpha_0,\theta_0)= S(\alpha_0,\theta_0)+\sup_{|\theta|\leq \theta_0}\ \sup_{\alphab\in\cB(\alpha_0)} \|V_{\betab+\alphab+\theta \bm 1}\|.
\end{equation}
Note that $\cB(\alpha_0)$ is the intersection of the real open $\ell^{\infty}$ ball of radius $\alpha_0$ with the hyperplane of equation $\alphab.\bm 1=0$.

Contrary to the simple estimate for $\chi_t(\alphab)$ in \Cref{prop_energycorr2}, a comparison of $\chi_t({\alphab})$ and $\chi_t({\alphab} + \theta{\bm 1})$ in the case where $\rho$ is proportional to $\e^{-\betab.{\bm E}}$ requires to control not only $\e^{+{\bm \alpha}. {\bm E}}\e^{+\i t H}\e^{-{\bm \alpha}. {\bm E}}$ but also $\e^{+{\frac12(\bm\beta-\bm \alpha)}. {\bm E}}\e^{+\i t H}\e^{-\frac12{(\bm\beta-\bm \alpha)}. {\bm E}}$. This explains the role of the second term in \eqref{Const-R2}, which is the same as $S(\alpha_0, \theta_0)$ with $V$ replaced by the deformed interaction observable $V_\betab$.

We then have the following result, which will be crucial in our statistical formulation of heat conservation. Its proof can be found in \Cref{subsec_proof_subsec_bounds}.
\begin{proposition} \label{cut}
Let $(\cH,\rho,\bm E,V)$ be a confined multisystem with $\rho$ multi-thermal at inverse temperatures $\betab$. Then for all ${\bm \alpha}\in\rr^\ell$ such that $\alphab.\bm 1=0$ and $\theta\in\rr$, 
\begin{equation} 
\chi_t({\bm \alpha})\, \e^{-|\theta|\, S_\betab(\norme{\alphab},\theta)}\leq 
 \chi_t({\alphab} + \theta{\bm 1}) \leq \chi_t({\bm \alpha})\, \e^{+|\theta|\, S_\betab(\norme{\alphab},\theta)} .
\end{equation}
\end{proposition}

\section{Multi-reservoir systems: the first and second laws} \label{sec_OpenQuantumSystems}
In this section we will study the joint distribution of the heat fluxes, as defined by the Two-Time Measurement picture, in each reservoir of a multi-reservoir systems. Our approach is to describe such systems as the thermodynamic limit of confined multisystems as described in \Cref{subsec_ConfinedSetup}. We will use the bounds proved in \Cref{subsec_bounds} and make assumptions of uniform boundedness of quantities such as $S(\alpha_0,\theta_0)$ or $S_\betab(\alpha_0,\theta_0)$. As we have argued in the introduction, we view these assumptions as ultraviolet conditions.

\subsection{Setup: multi-reservoir systems and thermodynamic limit} \label{sec-setup-general}
Our setting for a multi-reservoir system will be the following:
\begin{definition} A multi-reservoir system is a family of confined multisystems $(\cH\iL,\rho\iL, \bm E\iL,V\iL)_{L\in\nn}$ where $\bm E\iL$ is a vector with $\ell$ components and the index $\ell$ is independent of $L$. We say that the multi-reservoir system is in a multi-thermal state at inverse temperatures $\betab=(\beta_1,\ldots,\beta_\ell)$ if, for all $L$, $\rho\iL=\rho\iL_\betab$ as in \eqref{eq_GibbsState}. We say that the multi-reservoir system is time-reversal invariant if for every $L$ in $ \nn$, the confined multisystem $(\cH^{(L)},\rho\iL, \bm E^{(L)},V^{(L)})$, is time-reversal invariant.
\end{definition}
Note that the existence of an embedding of the $L$ system in the $L^\prime$-th system for $L<L^\prime$ plays no role in our arguments and will be neither assumed nor discussed. 

\begin{remark}
\label{remark_smallsys}
  A multi-reservoir system models a family of reservoirs, each of which is described through an approximating sequence of confined systems, and the interaction between those reservoirs. One may wonder what effect a ``small'' system, i.e.\ one whose Hilbert space dimension remains finite in the $L\to\infty$ limit,  can have on the thermodynamics of the system, in particular in the case where the reservoirs $\cR_j$ interact through this small system. Note that this will typically change  not only the state space of the considered random variables but also the reference probability distribution. In \Cref{subsec_smallsystem} we show that, in the large-time limit, this effect is essentially irrelevant.
\end{remark}

In the following, all quantities of interest related to the $L$-th confined system will be denoted with a superscript $(L)$ (e.g.\ $\pp_t^{(L)}$, $\chi_t^{(L)}$). We now state our minimal assumptions on the existence of the thermodynamic limit:
\begin{quote} \hypertarget{TL}{\textbf{Assumption \TL}: for $(t, {\bm \alpha}) \in \rr_+\times \i \rr^\ell$, the following limit exists and is finite:
\begin{equation} \label{eq_domcvgchitL}
\lim_{L\rightarrow \infty}\chi_t^{(L)}({\alphab})=\chi_t({\alphab}),
\end{equation}
and for all $t\in\rr_+$, $\i\rr^\ell\ni\bm \alpha\mapsto \chi_t(\bm \alpha)$ is continuous at $0$.}
\end{quote}
This is a standard assumption implying the existence of a Borel probability measure ${\pp}_t$ on $\rr^\ell$ such that for any bounded continuous function $f:\rr^\ell \rightarrow \cc$, $\lim_{L\rightarrow \infty}\int f({\bm\varphi})\,\d {\pp}_t^{(L)}({\bm\varphi})=\int f({\bm\varphi})\,\d {\pp}_t({\bm\varphi})$. We extend the definition of $\chi_t(\alphab)$ to any $\alphab\in\rr^\ell$ as an extended real number by setting $\chi_t(\alphab):=\int \e^{-t\alphab.{\bm \varphi}}\d\pp_t({\bm \varphi})$.\label{eq_defchitalphaRl}

We limit our study to bounded interactions, hence throughout the paper we assume
\begin{equation}\label{eq:bounded_V}
\sup_L \|V^{(L)}\|<\infty
\end{equation}
without further mention.

We will consider two regularity assumptions that strengthen this assumption. We recall that $S^{(L)}(\alpha_0,\theta_0)$ and $S^{(L)}_\betab (\alpha_0,\theta_0)$ are defined by relations \mref{Const-R} and \mref{Const-R2} respectively.
\begin{quote}\hypertarget{Blt}{\textbf{Assumption \Blt}: for some $\alpha_0>0$ and $\theta_0\geq \alpha_0$,
\begin{equation*}
S(\alpha_0,\theta_0){=}\sup_L S^{(L)}(\alpha_0,\theta_0) \ \mbox{is finite.}
\end{equation*}}
\end{quote}
\begin{quote}\hypertarget{Blbt}{
\textbf{Assumption \Blbt}: for some $\alpha_0>0$ and $\theta_0\geq \alpha_0$,
\begin{equation*}
S_\betab(\alpha_0,\theta_0){=}\sup_L S^{(L)}_\betab(\alpha_0,\theta_0) \ \mbox{is finite.}
\end{equation*}}
\end{quote}
Observe that \Blbt\ implies \Blt, but also that \Blt\ implies \Blbt\ (although with different values of $\alpha_0$ and $\theta_0$) for small enough values of the $\beta_j$, i.e.\ at high temperatures. 

We will use the following sets: 
\begin{gather} 
 \strip^{\alpha_0,\theta_0}=\{\alphab + s {\bm 1} \ \mbox{ s.t. }(\Re\alpha_j)_j\in\cB(\alpha_0), |\Re s|\leq \theta_0 \} \label{eq_defRalpha}\\
 \hull_{\betab}^{\alpha_0,\theta_0} = \big\{\alphab+t\betab\ \mbox{ s.t. } \alphab\in \strip^{\alpha_0,\theta_0}, t\in[0,1]\big\}.\label{eq_defCbetaalpha}
\end{gather}
Note that $\hull_{\betab}^{\alpha_0,\theta_0}$ is simply the convex hull of the sets $\strip^{\alpha_0,\theta_0}$ and $\betab+\strip^{\alpha_0,\theta_0}$. A schematic representation of the real part of these sets for $\ell=2$ is drawn in \Cref{fig:strip_and_hull}.

\begin{figure}
\begin{center}
\begin{tikzpicture}[node distance=2ex, scale=0.8]
\tikzmath{\a0=.3;\t0=1.4;\b1=1.5;\b2=3;} \tikzmath{\urc=5;\llc=2;} \fill[pattern=north west lines, pattern color=gray,opacity=1] (-\t0-\a0,-\t0+\a0)--(-\t0+\b1-\a0,-\t0+\a0+\b2)--(-\a0+\t0+\b1,\t0+\a0+\b2)--(\t0+\a0+\b1,\t0-\a0+\b2)--(\t0+\a0,\t0-\a0)--(-\t0+\a0,-\t0-\a0)-- cycle;
\draw[fill=gray, opacity=.6, draw=none] (-\t0-\a0,-\t0+\a0)--(-\a0+\t0,\t0+\a0)--(\t0+\a0,\t0-\a0)--(-\t0+\a0,-\t0-\a0)-- cycle;
\draw[opacity=.5] (-\t0+\b1-\a0,-\t0+\a0+\b2)--(-\a0+\t0+\b1,\t0+\a0+\b2)--(\t0+\a0+\b1,\t0-\a0+\b2)--(-\t0+\a0+\b1,-\t0-\a0+\b2)-- cycle;
\draw[->] (0,-\llc)--(0,\urc) node[above] {$\alpha_2$};
\draw[->] (-\llc,0)--(\urc,0) node[right] {$\alpha_1$};
\draw[color=gray,->] (-\llc,-\llc)--(\urc,\urc) node[right] {$\theta$};
\draw[color=gray,->] (\llc,-\llc)--(-\llc,\llc) node[right] {$\alphab$};
\node (beta) at (\b1,\b2) {\textbullet};
\node[right of=beta] {$\betab$};
\node (alpha0) at (\a0,-\a0) {\textbullet};
\node[right of=alpha0, node distance=5.5ex] {$(\alpha_0,-\alpha_0)$};
\node (theta0) at (\t0,\t0) {\textbullet};
\node[right of=theta0,node distance=4.5ex] {$(\theta_0,\theta_0)$};
\draw[fill=gray, opacity=.6, draw=none] (3.5,2)--(4,2)--(4,1.6) node[midway, right, color=black, opacity=1] {$\strip^{\alpha_0,\theta_0}$}-- (3.5,1.6)-- cycle;
\draw[pattern=north west lines, pattern color=gray,opacity=.8, draw=none] (3.5,1.4)--(4,1.4)--(4,1) node[midway,right, color=black, opacity=1] {$\hull_\betab^{\alpha_0,\theta_0}$}--(3.5,1)--cycle;
\end{tikzpicture}
\end{center}
\caption{\label{fig:strip_and_hull} Schematic representation of the domains $\strip^{\alpha_0,\theta_0}\cap \rr^\ell$ and $\hull_\betab^{\alpha_0,\theta_0}\cap \rr^\ell$ for $\ell=2$ with $\strip^{\alpha_0,\theta_0}$ and $\hull_\betab^{\alpha_0,\theta_0}$ defined in the text.}
\end{figure}

\begin{remark}
In Appendix \ref{sec_analyticapprox} we show that, starting from any multi-reservoir system with interaction $(V^{(L)})_{L\in\nn}$ satisfying the assumptions of uniform boundedness $\sup_{L} \|V^{(L)}\|<\infty$ and the local continuity in $0$: $\lim_{\|\alphab\|\to 0}\sup_{L} \|\e^{+\i\alphab.\bm E^{(L)}}V^{(L)}\e^{-\i\alphab.\bm E^{(L)}}-V^{(L)}\|=0$, we can find an (arbitrarily good) approximating sequence $(\tilde V^{(L)})_{L\in\nn}$ satisfying \Blbt\ for any $\betab$, $\alpha_0$ and $\theta_0$.
\end{remark}

A standard convergence theorem (Vitali's lemma \cite[Appendix B]{JOPP}) guarantees the existence and regularity of the limit $\lim_{L\rightarrow \infty}\chi_t^{(L)}({\alphab})$ for $\alphab \in \rr^\ell$. This is formulated in the next statement.

\begin{lemma}\label{lemma_VitaliChi}
Assume \TL\ and \Blt\ hold. Then for any $\alphab\in \strip^{\alpha_0,\theta_0}$, the limit \eqref{eq_domcvgchitL} exists. It is uniform in any compact subset of the strip $\strip^{\alpha_0,\theta_0}$, and defines an analytic function.
Moreover, for any non negative $t$ and any $\alphab$ in $\strip^{\alpha_0,\theta_0}$,
\begin{equation} \label{eq_defchit}
	\lim_{L\rightarrow \infty}\chi_t^{(L)}({\alphab})=\int \e^{-t\alphab.\bm{\varphi}} \,\d\pp_t(\bm\varphi).
\end{equation}
In addition, the family $(\pp_t)_t$ is exponentially tight.
\end{lemma}

\proof 
Under assumptions \TL\ and \Blt, the bound \eqref{eq_triangle} and \Cref{prop_energycorr2} allow us to apply Vitali's lemma on the set $\strip^{\alpha_0,\theta_0}$. The exponential tightness of $(\pp_t)_t$ follows from \Cref{prop_energycorr2} again, as
\begin{align*}
\pp_{t}(\norme{\varphib}\geq M) &\leq \e^{- t \alpha_0 M }\,\ee_{t}(\exp t \alpha_0 \norme{\varphib}) \leq \e^{-t\alpha_0 M } \sum_{\bm\epsilon\in\{-1,+1\}^\ell}\chi_t(\alpha_0 \bm \epsilon)
\end{align*}
which from \Blt\ is bounded by $2^\ell \e^{t 2 S(\alpha_0,0)} \e^{-t\alpha_0 M}$.
\qed

This shows that the distribution $\pp_{t}$ of $\phib$ (and therefore the distribution $\pp_{H_0,t}$ of $\phib_0$ under $\pp_t$) is light-tailed. This is non-trivial, as $\phib$ is not in general a bounded random variable; see \cite{BPR}. Note that condition \Blt was used.

It follows that, if we assume \TL\ and \Blt, then the thermodynamic limits of average fluxes exist: from relations \eqref{eq_DrvChit}, the limits
\[
\langle \Phi_j\rangle_t:=\lim_{L\rightarrow\infty} \ee_t\iL(\phi_j)=\lim_{L\rightarrow\infty}-\frac{1}{t}\frac{\partial}{\partial{\alpha_j}}\chi_{{t}}\iL(\alphab)_{|\alphab=\bm 0} = \lim_{L\rightarrow\infty} \langle \Phi_j^{(L)}\rangle_t
\]
exist and satisfy
\begin{equation} \label{eq_identitespostL}
	\langle \Phi_j\rangle_t=\ee_t(\phi_j)=-\frac{1}{t}\frac{\partial}{\partial{\alpha_j}}\chi_{{t}}(\alphab)_{|\alphab=\bm 0}.
\end{equation}

Transient fluctuations relations are immediately inherited from their confined version (\Cref{prop_EvansSearles}) whenever the limit exists. Finiteness of the limit is guaranteed by \Cref{lemma_VitaliChi} and H\"older's inequality. This is summarized in the following lemma:
\begin{lemma} \label{lemma_EvansSearles2}
For any multi-reservoir system which is time-reversal invariant, in a multi-thermal state at inverse temperatures $\betab$, and satisfies \TL, the quantity $\chi_t(\alphab)\in[0,+\infty]$ satisfies
\begin{equation} \label{eq_EvansSearles}
 \chi_t({\bm \alpha})=\chi_t(\betab -{\bm \alpha}) \ \mbox{ for any } \alphab\in \rr^\ell.
\end{equation}
If in addition \Blt\ holds then $\chi_t(\alphab)$ is finite for $\alphab$ in $\hull_\betab^{\alpha_0,\theta_0}$.
\end{lemma}

\subsection{Limiting generating functions and translational symmetry}
\label{sec-0-firstsecondlaw}

In this section we consider the large-time limit and we prove a translational symmetry of the limit function that will have important consequences on statistical refinements of the heat conservation. In the large-time limit, the relevant quantities are the rates $\frac1t\log\chi_t(\alphab)$. Such limits may not exist, and we therefore start by considering
\begin{gather} \label{eq_deflimsups}
\overline {\chi}_+({\alphab})=\limsup_{t\rightarrow \infty}\frac{1}{t}\log 
{\chi}_t({\alphab})
\end{gather}
for all $\alphab$ in $\rr^\ell$ (recall that by convention ${\chi}_t({\alphab})\in(0,+\infty]$ is well-defined for all $t$, see \Cref{eq_defchitalphaRl}). This allows us to discuss properties of $\overline \chi_+$ without worry. The function $\overline \chi_+$ vanishes at the origin, takes values in $[-\infty, \infty]$, and is convex. Similarly, whenever the limit is defined, we let 
\begin{equation} \label{eq_deflimits}
{\chi}_+({\alphab})=\lim_{t\rightarrow \infty}\frac{1}{t}\log 
{\chi}_t({\alphab})
\end{equation}
(we postpone the discussion regarding the domain of definition of $\chi_+$). Obviously, relations satisfied by $\overline \chi_+$ will carry over to $\chi_+$. The following theorem is obtained as an immediate consequence of the key bound of \Cref{cut}, and of \Cref{lemma_EvansSearles2}. 
\begin{theorem} \label{theo_AGMT}
Consider a multi-reservoir system in a multi-thermal state satisfying \TL. If 
 \Blbt\ holds, then for any $\alphab\in\cB(\alpha_0)$, $\theta\in[-\theta_0,\theta_0]$ we have 
 	\begin{equation} \label{eq_AGMTsymmetry}  
  		\overline {\chi}_+({\alphab} +\theta {\bm 1})= \overline\chi_+({\alphab}).
 	\end{equation}
  If the multi-reservoir system is time-reversal invariant, then for any $\alphab\in\rr^\ell$,
	\begin{equation} \label{eq_ESsymmetry}
  		\overline {\chi}_+(\betab-\alphab)= \overline\chi_+(\alphab).
 	\end{equation}
\end{theorem}

When the multisystem $(\cH,\rho,{\bm E},V)$ is such that $\rho$ is not necessarily multi-thermal, assuming that \Blt\ holds for some $\alpha_0$ and any $\theta_0$ leads to the translation symmetry \eqref{eq_AGMTsymmetry}.
\begin{theorem}\label{prop:trans_sym_nonthermal}
Consider a multisystem satisfying \TL\ and \Blt\ for some $\alpha_0>0$ and any $\theta_0\geq \alpha_0$. Then, for any $\alphab\in\cB(\alpha_0)$ and any $\theta\in\rr$,
\[\overline{\chi}_+(\alphab+\theta{\bm 1})=\overline{\chi}_+(\alphab).\]
\end{theorem}
\proof
\Cref{prop_energycorr2} implies $\overline{\chi}_+(\alphab+\theta \bm 1)<\infty$ for any $\alphab\in\cB(\alpha_0)$ and $\theta\in\rr$. As a superior limit of convex functions, $(\alphab,\theta)\to \overline{\chi}_+(\alphab+\theta\bm 1)$ is convex on $\cB(\alpha_0)\times\rr$. As a finite convex function, $\alphab\mapsto \overline{\chi}_+(\alphab)$
is continuous on $\cB(\alpha_0)$. 

Since \Blt\ holds for any $\theta_0$, $\sup_L S_{H_0}\iL(\theta)<\infty$ for any $\theta\in\rr$. Then \eqref{eq_BJPPP_bound} implies $\overline{\chi}_+(\theta\bm 1)=0$ for any $\theta\in\rr$. Assumption \Blt\ for any $\theta_0\in\rr$ and \Cref{lemma_VitaliChi} ensure $\chi_t(\alphab+\theta \bm 1)$ is finite for $(\alphab,\theta)\in\cB(\alpha_0)\times \rr$. Then, from H\"older's inequality, for any $t\in\rr$, any $\alphab\in\cB(\alpha_0)$ and any $\theta\in\rr$, for any $p>1$ small enough that $p\alphab\in\cB(\alpha_0)$ and $q>1$ such that $1/p+1/q=1$, 
\[\log\chi_t(\alphab+\theta\bm 1)\leq \frac1p\log\chi_t(p\alphab)+\frac1q\log\chi_t(q\theta\bm 1).\]
It follows that for $\alphab\in\cB(\alpha_0)$, $p>1$ small enough and any $\theta\in\rr$,
$\overline{\chi}_+(\alphab+\theta\bm 1)\leq \frac1p\, \overline{\chi}_+(p\alphab).$
Since $\alphab\mapsto \overline{\chi}_+(\alphab)$ is continuous on $\cB(\alpha_0)$, taking $p$ to $1$ leads to $\overline{\chi}_+(\alphab+\theta\bm 1)\leq \overline{\chi}_+(\alphab)$ for all $\theta$. By convexity of $\overline\chi_+$ one has equality.
\qed
\begin{remark}
We will not mention the assumption ``\Blt\ for some $\alpha_0>0$ and any $\theta_0\geq \alpha_0$'' again, but every time a result is a consequence of the symmetry \eqref{eq_AGMTsymmetry}, this assumption can be used instead of \Blbt\ for some $\alpha_0>0$ and some $\theta_0\geq \alpha_0$.
\end{remark}
\begin{remark}\label{rk:error_AGMT}
Assumption \Blbt\ is missing in \cite{AGMT}. In Step 5 of \cite[Section 4.4]{AGMT} the authors claim that \Blbt\ holds for any $\alpha_0$ and $\theta_0$ as long as \eqref{eq:bounded_V} holds. In \cite[Section 4.1]{BPR} a quasi-free fermion gas example is given where the validity of \Blbt\ depends on the ultraviolet regularity of the thermodynamic limit of $V^{(L)}$. Particularly, the example can be such that $\sup_L\|V\iL\|<\infty$ but there does not exist $\alpha_0>0$ and $\theta_0\geq \alpha_0$ such that \Blbt\ holds. Moreover, while \Blbt\ and relation \eqref{eq_BJPPP_bound} imply $\sup_{t\in\rr}\chi_t(-\theta_0 \bm 1)<\infty$, in this model \TL\ holds and $\sup_L\|V\iL\|<\infty$ but $\chi_t(-\theta\bm 1)=\infty$ for any $\theta>0$ and almost every time $t$.
\end{remark}

In the rest of this section we explore the consequences of the symmetries proved in \Cref{theo_AGMT} on the fluctuations of heat in the reservoirs.

For $\bm s\in \rr^\ell$, we define
\[\bar I(\bm s)=-\inf_{{\alphab}\in \rr^\ell}\big(\alphab. \bm s +\overline {\chi}_+({\alphab})\big)\in [0,+\infty].\]
This function $\bar I$ allows to control the large deviations of $\phib$: a standard argument using the Markov inequality and an optimization in the parameter $\alphab$ shows that for any $\mathbf{s}\in \rr^\ell$ and any $\epsilon>0$, one has
\begin{equation} \label{eq_upperlargedeviation}
  \limsup_{t\to\infty}\frac1t\log\pp_t(\|\phib-\mathbf{s}\|\leq \epsilon)\leq -\inf_{\|\mathbf{u}-\mathbf{s}\|\leq \epsilon}\overline I(\mathbf{u}).
\end{equation}
In the case where $\chi_+$ is assumed to exist, then the role of the associated function $I$ will be made more precise by the G\"artner--Ellis theorem.

The following result gives relevant relations satisfied by this function $\bar I$.
\begin{theorem}\label{thm:sym_to_barI}
Assume that \TL\ holds. Then,
\begin{enumerate}[label={(\roman*)}]
\item If \Blbt\ holds, then
\[\bar I(\bm s)\geq \theta_0\,|\bm{s}.\bm{1}|.\]
Hence, if for any $\theta_0$ \Blbt\ holds for some $\alpha_0$ , then $\bar I(\bm s)=+\infty$ if $\bm s . \bm 1\neq 0$.
\item If the system is time-reversal invariant, then
\[\bar I(\bm s)=\bar I(-\bm s)-\bm\beta.\bm s.\]
\end{enumerate}
\end{theorem}
\proof
For Item (i), $\overline{I}(\bm s)=-\inf_{\alphab\perp\bm 1}\inf_{\theta\in\rr} \alphab.\bm s + \theta \bm s.\bm 1+\overline{\chi}_+(\alphab+\theta \bm 1)\geq -\inf_{\alphab\in \cB(\alpha_0)}\inf_{|\theta|<\theta_0} \alphab.\bm s + \theta \bm s.\bm 1+\overline{\chi}_+(\alphab+\theta \bm 1)$. Theorem \ref{theo_AGMT} then implies $\overline{\chi}_+(\alphab+\theta \bm 1)=\overline{\chi}_+(\alphab)$ for any $\alphab\in\cB(\alpha_0)$ and $|\theta|<\theta_0$. Hence,
$\overline{I}(\bm s)\geq \theta_0|\bm s.1| -\inf_{\alphab\in \bm \cB(\alpha_0)} \big(\alphab.\bm s +\overline{\chi}_+(\alphab)\big)$ and $\overline{\chi}_+(0)=0$ implies that (i) holds. Item (ii) follows from \Cref{theo_AGMT}.
\qed

\Cref{thm:sym_to_barI} (i) implies the following heat conservation type result. \Cref{thm:sym_to_barI} (ii) is related with an expression of the fluctuation relations at the level of the rate function.
\begin{theorem} \label{theo_refinement1stlaw2}
Consider a multi-reservoir system in a multi-thermal state and assume that \TL\ and \Blbt\ hold. Then, for any Borel set $\bm B$ in $\rr^\ell$,
\begin{equation}\label{eq_TheoLDPDetailedEnergy}
\limsup_{t\rightarrow \infty}\frac{1}{t}\log {\pp_{t}}(\bm B)\leq-\theta_0\inf_{\bm s\in\bm B}|\bm s. \bm 1|.
\end{equation}
\end{theorem}
\proof  By \Cref{lemma_VitaliChi}, the family $(\pp_{t})_t$ is exponentially tight. Baldi's theorem (see e.g.\ Theorem 4.5.20 in \cite{DZ}) then implies that
\[\limsup_{t\rightarrow \infty}\frac{1}{t}\log {\pp_{t}}(\bm B) \leq\limsup_{t\rightarrow \infty}\frac{1}{t}\log {\pp_{t}}(\clos \bm B) \leq-\inf_{\bm s\in \clos \bm B} \bar I(\bm s)\]
with $\clos \bm B$ the closure of $\bm B$.
The result follows from \Cref{thm:sym_to_barI} (i) and the continuity of $\bm~s\mapsto~\bm s.\bm 1$.
\qed

\begin{remark} \label{remark_BJPPP}
 The above implies in particular bounds such as
 \[\limsup_{t\to\infty} \frac1t \log \pp_t(|\varphi_0|>\epsilon)\leq -\theta_0 \epsilon.\]
 Such bounds, however, can be derived by assuming a uniform bound on $S\iL_{H_0}(\theta_0)$ (defined in \eqref{eq_defSH0}), which is a condition weaker than \Blbt. See \cite{BJPPP,BPR} for this and related results on the $\pp_t$-distribution of $\phi_0$ .
\end{remark} 

\subsection{Law of large numbers, central limit theorem and large deviation principle} \label{subsec_refinements}

In this section we want to describe the behavior of the distribution $\pp_t$ of the heat fluxes $\phib$ beyond large deviation upper bounds. To this purpose, we will introduce additional assumptions in terms of large-time (LT) behavior of $\chi_t$, which allow to apply standard results in large deviation theory. Although proving them can be non-trivial, it is known that LT assumptions hold for relevant physical models (see \cite{JOPP} and \Cref{sec-models}). Our proofs make extensive use of the G\"{a}rtner-Ellis and Bryc theorems, which give information on the large time behavior of a random variable (heat fluxes in our case) from the regularity of a limiting object $\chi_+(\alphab)$. To this purpose, additional assumptions in terms of large-time limit are required. Our discussion consists of three levels of increasing precision, which correspond to a law of large numbers, a central limit theorem, and a large deviation principle. 

We assume \TL\ and start with the following assumption about the large-time behavior of the cumulant generating functions $\log \chi_t$.
\begin{quote} \hypertarget{LT}{
\textbf{Assumption \LT}: \TL\ holds and for some $\alpha_0>0$, the limit
	\begin{equation} \label{eq_defchiplus}
		{\chi}_+({\alphab})=\lim_{t\rightarrow \infty}\frac{1}{t}\log \chi_t({\alphab})		
	\end{equation}
	exists as a real number for any $\alphab\in\rr^\ell$ such that $\|\alphab\|<\alpha_0$.}
\end{quote}
\begin{remark}Assuming \Blt, if $\chi_+(\bm \alpha)$ exists as an extended real number, then by \Cref{prop_energycorr2}, $\|\bm \alpha\|\leq \alpha_0$ implies $\chi_+(\bm \alpha)<\infty$.
\end{remark}
\begin{remark}\label{rk:LT_extention}
Assuming \LT\ and \Blbt, \Cref{cut} implies that $\chi_+(\alphab+\theta\bm 1)$ exists as a limit and is finite for any $(\alphab,\theta)\in \cB(\alpha_0)\times [-\theta_0,\theta_0]$.
\end{remark}
The existence of $\chi_+$ together with additional regularity assumptions on $\chi_+$ will have various consequences on the behavior of the distributions $(\pp_t)_t$.

Assume first that \LT\ holds and $\chi_+$ is differentiable at the origin. We can then define for $j=1,\ldots,\ell$ the quantity
\begin{equation} \label{eq_fluxmoyens} 
\langle \Phi_j\rangle_+:=-\frac{\partial}{\partial \alpha_j}\,\chi_+(\alphab)_{|\alphab = \bm 0}. 	
\end{equation}
Lemma IV.6.3 in \cite{Ellis} and relations \eqref{eq_identitespostL} then show the existence of the average fluxes in the thermodynamic and large-time limits:
\begin{equation} \label{eq_formulePhijmoy}
	\langle \Phi_j\rangle_+=\lim_{t\rightarrow \infty}	\langle \Phi_j\rangle_t=\lim_{t\rightarrow \infty}\ee_t(\phi_j)=-\lim_{t\rightarrow \infty}\frac{1}{t}\frac{\partial}{\partial{\alpha_j}}\chi_{{t}}(\alphab)_{|\alphab=\bm 0}.
\end{equation} 
We denote by $\langle\bm{\Phi}\rangle_+$ the vector with components $\langle\Phi_j\rangle_+$.

The following is a formulation of the heat conservation at the level of the averages of fluxes $\phi_1,\ldots,\phi_\ell$. Note that, because we consider the distributions of the random variable $\phib$ with respect to different probability measures $\pp_t$, no almost-sure convergence statement can be given.
\begin{theorem} \label{theo_firstlaw_avg}
Consider a multi-reservoir system in a multi-thermal state and assume that \LT\ holds, and that $\chi_+$ is differentiable at the origin. Then $\langle\bm{\Phi}\rangle_+$ satisfies
\begin{equation} \label{first-law} 
\sum_j \langle \Phi_j\rangle_+=0,
\end{equation}
and for any small enough $\epsilon>0$, there exists a constant $C_{\bm E}(\epsilon)>0$ such that 
\begin{equation}\label{eq_cvgexponentielle}
\limsup_{t\to\infty} \frac1t \log \pp_t \big(\sup_{j=1,\ldots,\ell} |\phi_j - \langle \Phi_j\rangle_+|>\epsilon \big)\leq -C_{\bm E}(\epsilon).
\end{equation}
\end{theorem}
\proof \Cref{first-law} follows from \eqref{eq:bound_mean_cov} and \eqref{eq:bounded_V}. The second part of the statement, i.e.\ relation \eqref{eq_cvgexponentielle}, is similar to \Cref{eq_upperlargedeviation} and is a standard result (see e.g.\ the proof of Theorem II.6.3.\ in \cite{Ellis}). \qed

We now turn to a central limit theorem for the random variables $\phib$. For this assume in addition to \LT\ that there exists a complex neigbourhood $\mathcal O$ of the origin such that $\frac{1}{t}|\log \chi_t(\alphab)|$ is uniformly bounded for $t>1$ and $\alphab \in\mathcal O$ (remark that \Cref{prop_energycorr2} only gives a bound on $\log |\chi_t(\alphab)|$). An application of Vitali's lemma then shows that the limit ${\chi}_+({\alphab})$ can be extended to an analytic function on $\mathcal O$. We define for $j,k=1,\ldots,\ell$
\begin{equation}\label{eq_covarianceTCL}
D_{j,k}{:=}\frac{\partial^2}{\partial \alpha_j\partial \alpha_k} \chi_+({\alphab})_{|{\alphab}=\bm 0}= \lim_{t\to\infty} \frac1t \Big( \frac{\partial^2 \chi_t({\alphab})}{\partial \alpha_j\partial \alpha_k}\, - \frac{\partial \chi_t({\alphab})}{\partial \alpha_j} \frac{\partial\chi_t({\alphab})}{\partial \alpha_j}\Big)_{|{\alphab}=\bm 0}=\lim_{t\to\infty} t\, \mathrm{cov}_t(\phi_j,\phi_k)
\end{equation}
(again the second and third equalities are consequences of Vitali's lemma and relations \eqref{eq_DrvChit}). The $\ell\times \ell$ matrix $\bm D=(D_{j,k})_{j,k}$ is automatically real-symmetric positive semidefinite. It is not positive definite since \eqref{eq:bound_mean_cov} and \eqref{eq:bounded_V} imply $\sum_{j,k}D_{j,k}=0$. We then have

\begin{theorem} \label{theo_cvgTCL} 
Consider a multi-reservoir system in a multi-thermal state and assume that \LT\ holds and that there exists a complex neigbourhood $\mathcal O$ of the origin with
\[\sup_{t>1}\sup_{\alphab\in \mathcal O} \frac{1}{t}|\log \chi_t(\alphab)|<\infty.\]
Then the following convergence in distribution, with respect to the family $(\pp_t)_t$, holds:
\begin{equation}\label{eq_cvgTCL}
\sqrt t \big(\phib- \langle\bm \Phi\rangle_+\big) \underset{t\to\infty}{\longrightarrow} \mathcal N(0, \bm D).
\end{equation}
\end{theorem}

\proof  The convergence \eqref{eq_cvgTCL} follows from Bryc's theorem (see Appendix B in \cite{JOPP} for a multi-dimensional version of the lemma originally proven in \cite{Bryc}), and relations \eqref{eq_covarianceTCL}.
\qed

To formulate a large deviation principle, we need to introduce some additional notation. We simply apply the G\"artner-Ellis theorem, and follow the treatment of \cite{DZ}. We will assume \LT, and that $\chi_+(\alphab)$ is defined as an extended real number by \eqref{eq_defchiplus} for all $\alphab$ in $\rr^\ell$. We denote by $\mathcal D$ the set 
\[\mathcal D = \{\alphab \in\rr^\ell \ \mbox{ s.t. } \ \chi_+(\alphab)<+\infty\}.\]
Note that, under assumption \LT, we have $\strip^{\frac12\alpha_0,\frac12\alpha_0}\subset \mathcal D$.
We define for $\bm s$ in $\rr^\ell$
\[ I(\bm s)=-\inf_{{\alphab}\in \rr^\ell}\big(\alphab. \bm s + {\chi}_+({\alphab})\big)\in [0,+\infty]\]
and denote by $\mathcal F$ the set of $\bm s \in\rr^\ell$ such that there exists $\alphab \in\mathcal D$ with
\[\alphab .\bm s+ I(\bm s) <\alphab .\bm s' + I(\bm s') \ \mbox{ for all } \ \bm s'\in\rr^\ell\setminus\{{\bm s}\}.\]
We can now state our two theorems:
\begin{theorem}\label{thm:sym_to_I}
Assume that \LT\ holds and assume that \eqref{eq_defchiplus} defines $\chi_+(\alphab)$ as an extended real number for all $\alphab$ in $\rr^\ell$. Then,
\begin{enumerate}[label={(\roman*)}]
\item if \Blbt\ holds, then
\[I(\bm s)\geq \theta_0 \, |\bm s.\bm 1|.\]
Hence, if \Blbt\ holds for any $\theta_0$, $I(\bm s)=+\infty$ if $\bm s\notin \bm 1^\perp$.
\item if the system is time-reversal invariant, then
\[I(\bm s)=I(-\bm s)-\bm\beta.\bm s.\]
\end{enumerate}
\end{theorem}
\proof The theorem follows from $I=\bar I$ and \Cref{thm:sym_to_barI}.\qed

\begin{theorem} \label{theo_SecondLaw3}
	Consider a multi-thermal multi-reservoir system satisfying \LT, and assume that \eqref{eq_defchiplus} defines $\chi_+(\alphab)$ as an extended real number for all $\alphab$ in $\rr^\ell$. Then for any Borel set~$\bm B\subset \rr^\ell$:
\begin{equation} \label{eq_LDPphij}
-\inf_{\bm s\in {\rm int}(\bm B)\cap \mathcal F}I(\bm s)\leq \liminf_{t\rightarrow \infty}\frac{1}{t}\log {\pp}_t(\bm B)\leq \limsup_{t\rightarrow \infty}\frac{1}{t}\log {\pp}_t(\bm B)\leq
-\inf_{ \bm s\in {\rm cl}(\bm B)}I(\bm s).
\end{equation}

If we assume that \LT\ holds for all $\alpha_0$ and $\rr^\ell\ni\alpha\mapsto\chi_+(\alpha)$ is differentiable everywhere, then \eqref{eq_LDPphij} holds with $\mathcal F$ replaced with $\rr^\ell$.
\end{theorem}
\proof
Relation \eqref{eq_LDPphij} is obtained by a direct application of the G\"artner-Ellis theorem (see \cite{DZ}). If \LT\ holds for all $\alpha_0>0$ and $\chi_+$ is differentiable on $\rr^\ell$, then $\mathcal F=\rr^\ell$.
\hfill \qed

\subsection{Linear response theory} \label{subsec_linearresponse}
In this section we are interested in multi-reservoir systems which are near thermal equilibrium, in the sense that they are multi-thermal at an inverse temperature $\betab=(\beta_1,\ldots,\beta_\ell)$ such that for all $j$ one has $\beta_j\simeq \beta_\eq$ for some~$\beta_\eq>0$. The purpose of the linear response theory of multi-reservoir systems is to describe fluxes to the first order in the thermodynamical forces $\zeta_j=\beta_j-\beta_\eq$.

We suppose that for some $\beta_\eq>0$ and $\delta>0$ we have, for any $\betab$ with $\|\betab-\beta_\eq\bm 1\| <\delta$, a multi-reservoir system $(\cH\iL, \rho\iL, \bm E\iL,V\iL)_{L\in\nn}$ which is multi-thermal at inverse temperatures~$\betab$. We assume that $\cH\iL$, $\bm E\iL$, $V\iL$ do not depend on $\betab$, and that for any $\betab$ as above, the multi-reservoir system $(\cH\iL, \rho\iL, \bm E\iL,V\iL)_{L}$ satisfies assumption \LT\ (note that \LT\ depends on $\betab$ through the state $\rho$).

We let $\betab_{\eq}=(\beta_\eq, \ldots, \beta_\eq)=\beta_\eq\bm 1$ and $\betab=\betab_\eq+\bm\zeta$ so that $\bm\zeta=\bm 0$ corresponds to the equilibrium situation $\betab=\betab_\eq=\beta_\eq\bm 1$. We shall denote by e.g.\ $\overline\chi_t(\betab, {\alphab})$, $\chi_t(\betab, {\alphab})$ the functions $\overline\chi_t(\alphab)$, $\chi_t(\alphab)$ corresponding to the value $\betab$ of the inverse temperatures, and indicate the dependence of the currents on the thermodynamical forces, denoting them e.g.\ $\langle \Phi_{j}\rangle_{\zetab}$. Assuming that the functions $\zetab\mapsto \langle \Phi_j\rangle_{\zetab}$ are differentiable in $\bm \zeta$ at the origin, the kinetic transport coefficients are defined by 
\begin{equation}\label{eq_defLjk}
L_{j,k}=\frac{\partial \langle \Phi_k\rangle_{\bm\zeta}}{\partial{\zeta_j}}{}_{|{\bm \zeta}=\bm 0}.
\end{equation}
An immediate consequence of the mean heat conservation in the form \eqref{first-law} is $\sum_k L_{j,k}=0$ for all $j$. 
To discuss further properties of these coefficients, we strengthen \LT.
\begin{quote}\hypertarget{LR}{\textbf{Assumption} \LR: for some $\beta_\eq>0$ and $\delta>0$ there exist $\alpha_0>0$ and $\theta_0\geq \alpha_0+\beta_\eq$ such that for any $\betab$ with $\|\betab-\beta_\eq\bm 1\| <\delta$ the multi-reservoir system satisfies \Blbt\ and \LT, and for any $\alphab,\zetab\in \rr^\ell$ with $\norme{\alphab} <\alpha_0$ and $\norme{\zetab}<\delta$, the limit 
\begin{equation} \label{eq_defLR}
\chi_+({\betab_\eq+\zetab},{\alphab})=\lim_{t\rightarrow \infty}\frac{1}{t}\log \chi_t(\betab_\eq+\bm\zeta, {\alphab})
\end{equation}
exists and defines a map that is continuously differentiable in $\zetab$ and twice continuously differentiable in $\alphab$.}
\end{quote}

An immediate consequence of this assumption is that the transport coefficients \eqref{eq_defLjk} are well-defined. The next theorem shows that Onsager reciprocity relations can be easily proven as a simple consequence of the translational symmetry \eqref{eq_AGMTsymmetry} and symmetry \eqref{eq_ESsymmetry}, recovering the result of \cite{AGMT}.
\begin{theorem}\label{theo_onsager}
Consider a $\betab$-dependent multi-reservoir system as described above, satisfying \LR, and such that for any $\betab$ with $\|\betab-\beta_\eq \bm 1\|<\delta$ the multi-reservoir system is time-reversal invariant and multi-thermal at inverse temperature $\betab$. 
Then for every $j,k=1,\ldots,\ell$ we have
\[2 L_{j,k}=D_{j,k}(\betab_\eq),
\]
and in particular $L_{j,k}=L_{k,j}$.
\end{theorem}
\proof 
Using successively translational symmetry \eqref{eq_AGMTsymmetry} and symmetry \eqref{eq_ESsymmetry}, we have for $\|\alpha\|\leq \alpha_0$,
\begin{equation}
\overline {\chi}_+({\betab}_\eq +\zetab, {\alphab})=\overline{\chi}_+({\betab}_\eq +\bm\zeta, {\alphab}+\betab_\eq )=\overline{\chi}_+(\betab_\eq +\bm\zeta, \zetab -{\alphab}) .
\label{eq_PfOnsager}\end{equation}
Assumption \LR\ ensures that $\chi_+$ is defined at $(\betab_\eq +\zetab,\alphab)$ for $\norme{\betab}< \delta$, $\inf_{\theta\leq \theta_0}\norme{\alphab-\theta\mathbf{1}}< \alpha_0$, and is a $C^{1,2}$ function. By relation \eqref{eq_PfOnsager}, it satisfies
\begin{equation} \label{eq_PfOnsager2}
{\chi}_+(\betab_\eq+\bm\zeta, {\alphab})={\chi}_+(\betab_\eq +\bm\zeta, \bm\zeta-{\alphab}).
\end{equation}
Definitions \eqref{eq_fluxmoyens} and \eqref{eq_defLjk} imply that
\[L_{j,k}=-\frac{\partial^2}{\partial \zeta_j \partial \alpha_k}\chi_+(\betab_\eq+\bm \zeta,\alphab)_{|\bm{\zeta}=\alphab=\bm 0}.\]
Relation \eqref{eq_PfOnsager2} and a simple application of the chain rule give
\begin{equation*}
-\frac{\partial^2}{\partial{\zeta_j}\partial{\alpha_k}} \chi_+(\betab_\eq+\zetab,\alphab)_{|\bm{\zeta}=\alphab=\bm 0}= \frac12 \frac{\partial^2}{\partial{\alpha_j}\partial{\alpha_k}} \chi_+(\betab_\eq+\zetab,\alphab)_{|\bm{\zeta}=\alphab=\bm 0}.
\end{equation*}
Comparison with \eqref{eq_covarianceTCL} shows that $L_{j,k}=\frac12 D_{j,k}(\betab_\eq)$.
\qed

\begin{remark} The equalities $L_{j,k}=L_{k,j}$ are the Onsager reciprocity relations. The central limit theorem \eqref{eq_cvgTCL} and relation $L_{j,k}=2 D_{j,k}$ are the first part of the Fluctuation-Dissipation Theorem for multi-reservoir systems. The second part involves the validity of the Green-Kubo formulas for ${\bm L}$; their formulation and proof require introduction of the infinitely extended dynamical system describing multi-reservoir systems and we will not discuss them in this paper (see \cite{JOPP} for references). Here we can only prove
\[L_{j,k}=\lim_{t\to\infty} \lim_{L\to\infty}\frac1t \int_0^t\int_0^t \rho_{\betab_\eq}\iL \Big(\big(\Phi\iL_{j,s_1}-\rho\iL_{\betab_\eq}(\Phi\iL_{j,s_1})\big)\big(\Phi\iL_{k,s_2}-\rho\iL_{\betab_\eq}(\Phi\iL_{k,s_2})\big)\Big)\d s_1 \d s_2.\]
\end{remark}

\section{Examples}
\label{sec-models}
\subsection{Open quantum spin systems} \label{subsec:spinsys}

In this section we study a model of multi-reservoir system (in the sense of \Cref{sec_OpenQuantumSystems}) defined from quantum spin systems.

Let $\mathcal G=\zz^d$, and fix a finite partition ${\mathcal G}={\mathcal G}_1\cup\ldots\cup {\mathcal G}_\ell$ of $\mathcal G$ into infinite sets. For each $j$, let $\partial \mathcal G_j\subset \mathcal G_j$ be the boundary of $\mathcal G_j$, i.e.\ the set of elements $x$ of $\mathcal G_j$ that have at least one neighboring vertex not in $\mathcal G_j$. To each vertex $x\in \mathcal G$ we associate a copy of $\cc^k$. Let $S$ be a self adjoint operator from $(\cc^k)^{\otimes(2d+1)}$ to itself. For any $x\in \mathcal G$ let $S_x$ be a copy of $S$ that acts non trivially only on the Hilbert space associated to $x$ and its neighboring vertices.

We then denote $\mathcal G\iL=\{-L,\ldots,+L\}^d$ and ${\mathcal G}_j\iL={\mathcal G}_j\cap \mathcal G\iL$ the restriction of $\mathcal G_j$ to $\mathcal G\iL$ for any $L\in \nn$ and assume for simplicity that ${\mathcal G}_j\cap \mathcal G\iL\neq\emptyset$ for any $L$ and~$j$. We similarly define $\partial \mathcal G_j\iL$. We then define
\begin{equation*}
H_j\iL= \sum_{x\in {\mathcal G}_j\iL\setminus \partial \mathcal G_j\iL} S_x
\end{equation*}
and
\begin{equation*}
  V\iL =\sum_{x\in \cup_{j=1}^\ell \partial \mathcal G_j\iL} J(x) S_x
\end{equation*}
with $x\mapsto J(x)$ a real function such that $\sum_{x\in \cup_{j=1}^\ell \partial \mathcal G_j}|J(x)|<\infty$. This assumption ensures that $V\iL$ stays bounded in the thermodynamic limit.

We recall that $\bm E^{(L)}=(H_1^{(L)}, \ldots, H_\ell^{(L)})$ and fix some $\betab=(\beta_1,\ldots,\beta_\ell)$. We assume that the system is multi-thermal at temperatures $\betab$, i.e.\ we consider the density matrix $\rho\iL=\e^{-\betab.\bm E^{(L)}}/\tr(\e^{-\betab.\bm E^{(L)}})$. This defines a multi-reservoir system $(\cH\iL,\rho\iL, \bm E\iL,V\iL)_{L\in\nn}$ in the sense of Section \ref{sec-setup-general}. 

The following result shows that \Cref{theo_refinement1stlaw2} applies to this example. It is proven in \Cref{sec:proofs_for_section_yyy}.
\begin{theorem}\label{thm-TDL spins}
  Suppose that $\lim_L \tr(\rho\iL A)=\rho(A)$ exists for any $A\in\mathcal O$. Then \TL\ holds, and one has \Blt\ for $\alpha_0$ and $\theta_0$ small enough.
\end{theorem}

\begin{remarks}\hfill
\begin{itemize}[noitemsep,topsep=0pt]
 \item \Cref{thm-TDL spins} can be generalized by looking at convergent subsequences of $\rho^{(L)}$. Then for each of these subsequences assumption \TL\ is true. This extension is relevant when the uniqueness of the limit $(\beta_j,\tau_j)$-KMS state for each part $j$ is not guaranteed.
 \item If $d=1$, one can prove, adapting the results of Araki \cite{Araki1D}, that \Blbt\ holds for any $\alpha_0,\theta_0\in\rr_+$.
\item Actually \TL\ and \Blt\ hold for a much more general set of spin models. The proof in \Cref{sec:proofs_for_section_yyy} can be easily extended using the appropriate adaptation of the proof of \cite[Theorem 6.2.4]{BR2}.
\end{itemize}
\end{remarks}

\subsection{Spin-fermion model} \label{subsec_spinfermion}
We now turn to the spin-fermion model, which describes a two-level atom interacting with $\ell$ independent free Fermi gas reservoirs. We will define a multi-reservoir system (in the sense of Section \ref{sec-setup-general}) through finite-dimensional approximations of the reservoirs, as in Example 5.3 of \cite{AJPP1}. We use freely standard notation which can be found in e.g.\ \cite{BR2}.

To describe the $\ell$ reservoirs, we consider for $j=1,\ldots,\ell$ a Hilbert space $\ch_j=L^2(\rr_+,\d x;{\mathfrak H}_j)$ for some auxiliary Hilbert space ${\mathfrak H}_j$, $h_j$ the operator of multiplication by the variable~$x~\in~\rr_+$, and a vector $v_j\in \ch_j$ which we call a form factor. The Hilbert spaces $\{\mathfrak H_j\}$ encode the non-energetic degrees of freedom of each particle, $h_j$ is the one-particle energy operator and the $v_j$'s will be the form factors of the interaction of  each reservoir with the spin. We denote by $a^*_j, a_j, \varphi_j$ the operators associated with the free Fermi gas $\Gamma_{\mathfrak f}(\ch_j)$. The small system (see Section \ref{subsec_smallsystem}) is described by the state space $\cH_\cS=\cc^2$ and the Hamiltonian $H_\cS~=~\sigma^{(3)}=\begin{pmatrix*}[r] 1 & 0 \\ 0 & -1\end{pmatrix*}$. Since, by the discussion in Section \ref{subsec_smallsystem}, the state of the small system has no influence on the large-time behavior, for notational simplicity we assume the intial state on $\cS$ is $\rho_\cS=\frac12 \id$.

The spin-fermion model is usually defined and studied without tight binding assumption, i.e.\ for an unbounded one-particle Hamiltonian, hence our choice of $h_j$. Tight binding models have a natural ultraviolet sharp cutoff induced by the lattice approximation, so ultraviolet conditions are always satisfied as we will see in the electronic black box model of \Cref{sec-EBB}. A continuous space model provides a non-trivial example in the sense that our conditions are satisfied only for some choices of the form factors.

To fit our previous framework, we need to assume the model can be obtained as limit of confined approximations.  This is summarized in condition \SFTL\ below. This assumption can easily be proved for relevant choices of Hilbert spaces $\{\mathfrak H_j\}$. Proofs will also be given in terms of confined approximations. However note that conditions \SFUV\ and \SFtwo\ and \SFthree\ are indeed conditions on the form factor that can be verified directly in the infinite dimensional model.

\begin{quote} \hypertarget{SFTL}{\textbf{Assumption} \SFTL: for every $j=1,\ldots, \ell$ and $L\in \nn$, there exist $\ch_j\iL, h_j\iL, v_j\iL$ with
\begin{enumerate}[noitemsep,topsep=0pt]
\item $\ch_j\iL$ is a finite-dimensional subspace of $\ch_j$ with $\ch_j\iL\subset\ch_{j}^{(L+1)}$ and $\overline{\bigcup_L\ch_j\iL}=\ch_j$,
\item $h_j\iL$ is a definite-positive operator on $\ch_j\iL$ and, if we extend canonically $h_j\iL$ to $\ch_j$, then $h_j\iL\to h_j$ in the strong resolvent sense as $L\to\infty$,
\item $v_j\iL$ is an element of $\ch_j\iL$ such that $v_j\iL \to v_j$ in $\ch_j$.
\end{enumerate}}
\end{quote}

We can then define $\cH\iL = \cH_\cS \otimes \bigotimes_{j=1}^\ell \Gamma_{\mathfrak f}(\ch_j\iL)$. We let $H_j\iL = \dGa(h_j\iL)$, $H_0\iL=\sum_{j=1}^\ell H_j\iL$ and $V\iL=\sum_{j=1}^\ell \sigma^{(1)}\otimes \varphi_j(v_j\iL)$, where all operators are extended canonically to $\cH\iL$. We also define, for fixed $\betab=(\beta_1,\ldots,\beta_\ell)$ by $\rho\iL$ the state $\rho_\cS\otimes \bigotimes_{j=1}^\ell \e^{-\beta_j H_j\iL}\!\!/Z_j\iL$ with $Z_j\iL=\tr (\e^{-\beta_j H_j\iL})$.

Under these assumption we have the following, which we prove in \Cref{sec:proofs_for_section_yyy}.
\begin{proposition} \label{prop_TLBspinfermion}
Assume \SFTL\ is satisfied. Then, for any $\lambda>0$ and $\betab$, assumption \TL\ holds. 
\end{proposition}

Conditions \Blt\ and \Blbt\ will be guaranteed by the following assumption defined for any $\gamma_0\in\rr_+$. Note that this assumption corresponds to the form factor superexponential decay in high frequencies.
\begin{quote}\hypertarget{SFUV}{{\bf Assumption} \SFUV: for all $j=1,\ldots, \ell$ one has $v_j \in \Dom(e^{\frac12\gamma_0 h_j})$.} 
\end{quote}

Again the following is proven in \Cref{sec:proofs_for_section_yyy}:
\begin{proposition} \label{prop_SFB}
Assume \SFTL\ and \SFUV. Then assumption \Blt\ is satisfied for all $\alpha_0$ and $\theta_0$ such that $\alpha_0\leq\theta_0\leq \gamma_0$, and assumption \Blbt\ is satisfied for all $\alpha_0$, $\theta_0$ and $\betab$ such that $\alpha_0\leq\theta_0\leq \gamma_0-\|\betab\|$.
\end{proposition}

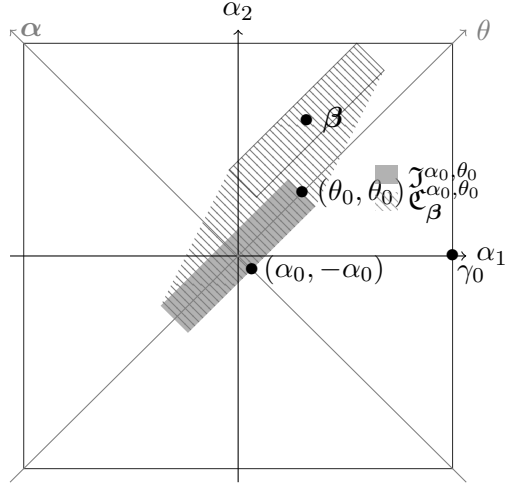
\begin{figure}
\begin{center}
\begin{tikzpicture}[node distance=2ex, scale=0.6]
\tikzmath{\a0=.3;\t0=1.4;\b1=1.5;\b2=3;} \tikzmath{\g0=\t0+\b2+\a0;}
\tikzmath{\urc=5;\llc=5;} \tikzmath{\lx=3;}\fill[pattern=north west lines, pattern color=gray,opacity=1] (-\t0-\a0,-\t0+\a0)--(-\t0+\b1-\a0,-\t0+\a0+\b2)--(-\a0+\t0+\b1,\t0+\a0+\b2)--(\t0+\a0+\b1,\t0-\a0+\b2)--(\t0+\a0,\t0-\a0)--(-\t0+\a0,-\t0-\a0)-- cycle;
\draw[fill=gray, opacity=.6, draw=none] (-\t0-\a0,-\t0+\a0)--(-\a0+\t0,\t0+\a0)--(\t0+\a0,\t0-\a0)--(-\t0+\a0,-\t0-\a0)-- cycle;
\draw[opacity=.5] (-\t0+\b1-\a0,-\t0+\a0+\b2)--(-\a0+\t0+\b1,\t0+\a0+\b2)--(\t0+\a0+\b1,\t0-\a0+\b2)--(-\t0+\a0+\b1,-\t0-\a0+\b2)-- cycle;
\draw[color=black, opacity=.8] (-\g0,\g0) rectangle (\g0,-\g0);
\draw[->] (0,-\llc)--(0,\urc) node[above] {$\alpha_2$};
\draw[->] (-\llc,0)--(\urc,0) node[right] {$\alpha_1$};
\draw[color=gray,->] (-\llc,-\llc)--(\urc,\urc) node[right] {$\theta$};
\draw[color=gray,->] (\llc,-\llc)--(-\llc,\llc) node[right] {$\alphab$};
\node (beta) at (\b1,\b2) {\textbullet};
\node[right of=beta] {$\betab$};
\node (alpha0) at (\a0,-\a0) {\textbullet};
\node[right of=alpha0, node distance=5.5ex] {$(\alpha_0,-\alpha_0)$};
\node (theta0) at (\t0,\t0) {\textbullet};
\node[right of=theta0,node distance=4.5ex] {$(\theta_0,\theta_0)$};
\node (gamma0) at (\g0,0) {\textbullet};
\node[below right of=gamma0] {$\gamma_0$};
\draw[fill=gray, opacity=.6, draw=none] (\lx,2)--(\lx+.5,2)--(\lx+.5,1.6) node[midway, right, color=black, opacity=1] {$\strip^{\alpha_0,\theta_0}$}-- (\lx,1.6)-- cycle;
\draw[pattern=north west lines, pattern color=gray,opacity=.8, draw=none] (\lx,1.4)--(\lx+.5,1.4)--(\lx+.5,1) node[midway,right, color=black, opacity=1] {$\hull_\betab^{\alpha_0,\theta_0}$}--(\lx,1)--cycle;
\end{tikzpicture}
\end{center}
\caption{\label{fig:strip_and_hull_SF} Schematic representation of the domains $\strip^{\alpha_0,\theta_0}\cap \rr^\ell$ and $\hull_\betab^{\alpha_0,\theta_0}\cap \rr^\ell$ for $\ell=2$ compared to the constant $\gamma_0$ in the spin-fermion model.}
\end{figure}

A schematic representation of the real part of $\strip^{\alpha_0,\theta_0}$ and $\hull_\betab^{\alpha_0,\theta_0}$ with respect to $\gamma_0$ for $\ell=2$ is presented in \Cref{fig:strip_and_hull_SF}.

For the large-time limit, we make additional technical assumptions. We extend the form factors $v_j$ to functions $\tilde v_j$ on $\rr$ by setting $\tilde v_j(x)=v_j(|x|)$.
The fist is a technical assuption, which is verified for physical choices of $v$

\begin{quote}\hypertarget{SFtwo}
{\textbf{Assumption} \SFtwo: there exists $\delta >0$ such that for every $j=1,\ldots,\ell$, the functions $\tilde v_j$ extend to analytic functions on the strip $|\Im z|<\delta$ satisfying
\[
\sup_{|y|<\delta}\int_{\rr}\e^{-\beta_j x}\|\tilde v_{j}(x+\i y)\|_{{\mathfrak H}_j}^2 \d x <\infty.
\]}
\end{quote}

The next assumption states that the small system is effectively coupled to the reservoir at 
 the Bohr frequency $(+1)-(-1)=2$ of the $2$-level system $\cS$. 
\begin{quote}
\textbf{Assumption} \SFthree: for all $j=1,\ldots,\ell$ one has $\|v_j(2)\|_{{\mathfrak H}_j}>0$.
\end{quote}

Remark also that \SFtwo\ implies \SFUV\ for any $\gamma_0\leq\min_{j=1,\ldots,\ell}\beta_j=:\beta_*$. Indeed it implies that for $y=0$ and any $\gamma_0\leq\beta_*$, $\int_{-\infty}^0\e^{-\gamma_0 x} \|\tilde{v}_j(x)\|_{\mathfrak{H}_j}^2\d x<\infty$. A simple change of variable $x\to-x$ leads to $\|e^{\frac12\gamma_0 h_j}v_j\|^2<\infty$ for any $\gamma_0\leq\beta_*$. Nevertheless, for clarity's sake, we keep referring to both assumptions independently.

In the sequel, we denote by e.g.\ $\chi_t^{(\lambda)}$ the generating function $\chi_t$ obtained in the thermodynamic limit for the system with coupling constant $\lambda$. 
\begin{theorem} \label{mainthm}
Suppose that \SFtwo\ and \SFthree\ hold. Let $\betab\in(0,+\infty)^\ell$ and $\epsilon,\delta >0$ be given. Then there exist $\Lambda>0$ and an open set ${\cal O}$ in $\cc^{2\ell}$ that contains $\big(\betab+(-\epsilon,+\epsilon)^\ell\big)\times (-\delta,+\delta)^\ell$ such that the following holds:
 \begin{enumerate}[noitemsep,topsep=0pt]
  \item For any $\lambda$ with $|\lambda|<\Lambda$ there exists $t_\lambda >0$
   such that for $t>t_\lambda$ the function $(\betab', \alphab)\mapsto \log\chi_{t}^{(\lambda)}(\betab', \alphab)$ has an analytic continuation to ${\cal O}$ that satisfies
   \begin{equation} \label{eq_SFone}
    \sup _{t >t_\lambda} \sup_{(\betab',\alphab)\in {\cal O}}\frac{1}{t}|\log\chi_t^{(\lambda)}(\betab',\alphab)|<\infty.
   \end{equation}
  \item For any $\lambda$ with $|\lambda|<\Lambda$, the limit
   \begin{equation} \label{eq_SFtwo}
    \chi_+^{(\lambda)}(\betab',\alphab)=\lim_{t\rightarrow \infty}\frac{1}{t}\log\chi_t^{(\lambda)}(\betab',\alphab)
   \end{equation}
   exists and defines a real analytic function on $\big(\betab+(-\epsilon,+\epsilon)^\ell\big)\times (-\delta,+\delta)^\ell$. 
 \end{enumerate}
\end{theorem}

The above statement can be proven using the spectral scheme described in \cite{JOPP}[Sect 5.5] combined with the remarks for the generalization to the multiparameter case in \cite{JOPP}[Sect 6.5]. In \cite{DR} a similar analysis of the more involved spin-boson model has been done.

Clearly by \Cref{prop_TLBspinfermion,prop_SFB} and \Cref{mainthm}, a spin-fermion model with \SFTL, \SFUV, \SFtwo\ and \SFthree\ satisfies assumptions \TL, \Blbt, \LT\ with regularity of $\chi_+$ in a complex neighbourhood of the origin, for any $\alpha_0$ and $\theta_0$ with $\alpha_0+\theta_0\leq\gamma_0$, and in addition \LR\ for any $\beta_\eq>0$ and $0<\delta<\beta_\eq$, as soon as $\lambda$ is chosen small enough. Therefore, \Cref{theo_refinement1stlaw2,theo_firstlaw_avg,theo_cvgTCL} apply.

Lastly, the present spin-fermion model is time-reversal invariant if all functions $v_j$ are real: if $C_\cS$ is the complex conjugation in the canonical basis of $\cH_\cS$ and $c_j\iL$ the complex conjugation of $\ch_j$ then $C\iL=C_\cS\otimes \bigotimes_{j=1}^\ell \Gamma(c_j\iL)$ is a time-reversal of the $L$-th confined model. If this is assumed in addition to \SFTL, \SFUV, \SFtwo\ and \SFthree, then \Cref{theo_onsager} holds as well.

\subsection{Electronic Black Box model} \label{sec-EBB}
The electronic black box (EBB) model is a basic paradigm in the study of coherent transport in electronic systems in mesoscopic physics (see \cite{AJPP1, JOPP} for more references). It consists of $\ell$ infinitely extended leads exchanging quasi-free fermionic particles through a finite system $\sys$. A distinctive feature of this model is the tight-binding approximation. This discretization of space leads to a sharp ultraviolet cut off. The energy of each fermion is bounded, hence the one particle Hamiltonian is bounded and all our ultraviolet regularity assumptions are verified for any parameter.

The Hilbert space describing the $L$-th confined system is $\cH\iL=\Gamma_{\mathfrak f}(\ch\iL)$ where $\ch\iL=\ch_\sys\oplus \bigoplus_{j=1}^\ell \ch_j\iL$. For simplicity we consider $\ch_\sys=\cc$ and $\ch_j\iL=\ell^2(\{0,\ldots,L\})$. We use the canonical embedding to identify each of $\ch_\sys$, $\ch_j\iL$ with a subspace of $\ch\iL$. The associated Hamiltonians are
\[ H_\sys= \dGa (h_\sys), \quad H_j\iL=\dGa (h_j\iL),\quad H\iL=\dGa(h\iL)\]
where $h_\sys=\epsilon_0$, $h_j\iL=-\frac12 \Delta_j\iL$ for $\Delta_j\iL$ the discrete Laplacian with Dirichlet boundary conditions $u_{-1}=u_{L+1}=0$, and $h\iL=h_\sys+\sum_j h_j\iL+ \lambda v\iL$ for $v\iL = \sum_j\big(|\chi\rangle \langle \delta\iL_j| + |\delta\iL_j\rangle \langle \chi|\big)$ where $\chi=1\in\cc$ and $\delta\iL_j=(1,0,\ldots)\in\ch_j$. Remark that for all $j$, $\sup_L \|h_j\iL\|<\infty$. The quantity $\lambda>0$ is a coupling constant and we denote $h_0\iL=h_\sys+\sum_j h_j\iL$. We assume that the initial state of the reservoirs is multi-thermal, and the initial state of $\sys$ is chaotic, i.e.\ $\rho\iL=\frac12 \id\otimes\bigotimes_{j=1}^\ell \e^{-\beta_j H_j}/Z_j\iL$. Once again we discard the heat variation of the small system (see \Cref{subsec_smallsystem}) and consider the generating function $\chi_{t} \iL$ for the heat variation of the reservoirs only.

We denote by $\eb\iL=(h_1\iL,\ldots,h_\ell\iL)$. Using the general identity $\tr \,\Gamma(A)= \det (\id + A)$ (see \cite[Lemma 6.1]{JOPP}) we have immediately
\begin{equation}
 \chi_{t}\iL(\alphab)=\frac{\det \big(\id+\e^{-\i t h\iL} \e^{(\alphab-\betab)\cdot\eb\iL}\e^{+\i th\iL} \e^{-\alphab\cdot \eb\iL} \big)}{\det(\id + \e^{-\betab\cdot \eb\iL})}=\frac{\det\big(\id + \e^{-\betab.\eb\iL} \e^{\i t h\iL_{\alphab}}\e^{-\i t h\iL}\big)}{\det\big(\id + \e^{-\betab.\eb\iL}\big)}.
\end{equation}
with $h_{\alphab}=h_0+\lambda v_{\alphab}$ for 
\[v_{\alphab}\iL=\sum_j \big(|\chi\rangle \langle \e^{-\alpha_j h_j}\delta\iL_j| + |\e^{\alpha_j h_j}\delta\iL_j\rangle \langle \chi|\big).\]
Remark that the $\|{\e^{\alpha h_j\iL} \delta\iL_j}\|$ are bounded uniformly in $L\in\nn$ and in $\alphab$ for $\norme{\Re\alpha}<\alpha_0$, for any $\alpha_0$ (in comparison with the general spin-fermion model, \SFUV\ holds for any $\gamma_0$ in the present model thanks to the boundedness of the discrete Laplacian). This shows:
\begin{proposition}\label{prop_BBEBB}
For any $\lambda>0$ and $\betab\in(0,\infty)^\ell$, assumption \TL\ holds, and assumption \Blbt\ is satisfied for all $\alpha_0,\theta_0 \in [0,+\infty)$.
\end{proposition}
 
We now wish to compute the quantity $\chi_+$; following the results of \Cref{subsec_smallsystem} we restrict to the case $\alpha_\sys=0$. Computing $\frac{\partial}{\partial\alpha_j}\log \chi_{t}^{(L)}(0,\alphab)$ for $j=1,\ldots,\ell$ one shows that (see Section 6.6.6 of \cite{JOPP} for details on all computations described in this section):
\begin{equation}
\frac1t\log \chi_{t}^{(L)}(0,\alphab)
=- \int_0^1 \!\int_0^1 \tr \big( ( \id +\e^{(\betab- u\alphab).\eb^{(L)}_{(1-r)t}}\, \e^{u\alphab.\eb^{(L)}_{-rt}} )^{-1} \, \i [ h\iL, \alphab . \eb\iL]\big)  \,\d r \,\d u
\label{eq-EBB-BC}
\end{equation}
with $\eb\iL_t=(h_{1,t}\iL,\ldots,h_{\ell,t}\iL)$ for $h_{j,t}\iL=\e^{-\i t h\iL} h_j\iL \e^{+\i t h\iL}$. The above formula remains valid after the thermodynamic limit, with operators $h_0$, $h$ (defined similarly to $h_0\iL$, $h\iL$ with a discrete Laplace operator on $\ell^2(\nn)$ satisfying Dirichlet conditions at $0$) replacing $h_0\iL$, $h\iL$. To discuss the large-time limit, we make the following additional assumption:
\begin{quote} 
\hypertarget{EBBTL}{{\textbf{Assumption} \EBBTL} The one-particle coupled Hamiltonian $h$ has only absolutely continous spectrum.}
\end{quote} 
Precise assumptions leading to \EBBTL\ (typically for any small enough $\lambda$) can be given, see e.g.\ Theorem 6.2 of \cite{AJPP1}. Next proposition follows from a straightforward adaptation of the arguments found in Section 6.6.6 of \cite{JOPP}. Provided \EBBTL\ holds, the following wave operators exist as strong limits:
$$W_\pm=\lim_{t\to\pm\infty}e^{ith}e^{-ith_0}\one_{\mathcal R}$$
with $h_0=h_{\mathcal S}+\sum_j h_j$ and $\one_{\mathcal R}$ the orthogonal projector onto $\bigoplus_j \mathfrak h_j$. Then we can give a useful expression for the limit cumulant generating function.
\begin{proposition} \label{prop_EBBLT}
 Assume \EBBTL, then the electronic black box model satisfies \LT\ for any $\alpha_0$. Let $S=W^*_+W_-$ be the scattering matrix and $T=W^*_- \, \i [ h, \alphab\cdot \eb] W_-$. We have
\begin{equation} \label{eq_ebb model}
 \chi_{+}(\alphab)=-\int_0^1 \tr_{\cH_\cR}\big((\id_\cR +S^*\e^{(\betab- u\alphab) \cdot{\eb}}S \,\e^{u\alphab.{\eb}})^{-1} \,T\big) \, \d u. 
\end{equation}
and the electronic black box model satisfies \LR\ for any $\alpha_0$ and $\delta$.
\end{proposition} 

Noticing that $T$ is finite rank, \Cref{eq_ebb model} yields that \LR\ holds for all $\alpha_0$ and $\delta$ and that there exists a complex neigbourhood $\mathcal O$ of the origin such that $\frac{1}{t}|\log \chi_t(\alphab)|$ is uniformly bounded for $t>1$ and $\alphab \in\mathcal O$. Therefore, under assumption \EBBTL, \Cref{theo_refinement1stlaw2,theo_firstlaw_avg,theo_cvgTCL,theo_SecondLaw3} hold for the electronic black box model.

As for time-reversal invariance, if the state $\rho_\cS\iL$ is Gibbs then the Electronic Black Box model is automatically TRI: if $c_\cS$ and $c_1\iL,\ldots,c_\ell\iL$ are the complex conjugations of $\ch_\cS$ of $\ch_j\iL$ then the operator $C\iL=\Gamma(c_\cS)\oplus \bigoplus_{j=1}^\ell \Gamma(c_j\iL)$ is a time-reversal invariant of the $L$-th confined system, and \Cref{theo_onsager} holds. Note that, denoting $J$ the complex conjugation on $\ch_s\oplus \ch_1\ldots\ch_\ell$, one has $Jh_0=h_0J$ and $Jh=hJ$, therefore $S^*=JSJ$.

One can also write $\chi_{+} (\alphab)$ in the momentum representation. By discrete Fourier transform, one can identify $\ch_j$ with $L^2([0, \pi], \d \xi_j)$ and $h_j$ with the multiplication operator by $\epsilon(\xi_j)=1-\cos\xi_j$. Moreover we can identify $\ch_\cR= \sum_{j=1}^\ell L^2([0, \pi], \d \xi_j)$ with $L^2([0, \pi], \d \xi) \otimes \cc^\ell$.
In this representation the scattering matrix acts as the operator of multiplication by a unitary $\ell\times \ell$ matrix $S(\xi)=\big(s_{j,j'}(\mathbf{\xi})\big)_{j,j'}$, and $T$ has an integral kernel $T(\xi^\prime,\xi)$ (see Equation 6.18 in \cite{JOPP}) with diagonal given by
\[
T(\xi,\xi)=\frac{\epsilon^\prime(\xi)}{2\pi} \big(S^*(\xi) K(\alphab,\xi)S(\xi)-K(\alphab,\xi)\big)
\] 
where $K(\alphab,\xi)$ is the diagonal matrix on $\cc^\ell$ with $j$th coefficient $K(\alphab,\xi)_j= - \alpha_j \epsilon(\xi)$.

Again following the arguments in Section 6.6.6 of \cite{JOPP} we find:
\begin{equation*}
\label{eq_ebb model2}
\chi_{+}(\alphab)=\frac1{2\pi} \int_0^{\pi}  
\log\frac{ \det (\id +S^*(\xi)\,\e^{-K(\betab- \alphab,\,\xi)}\, S(\xi) \e^{-K(\alphab,\,\xi)} )}{\det (\id+ \e^{-K(\betab,\,\xi)})}\,\epsilon^\prime(\xi)\,\d \xi.
\end{equation*}
 From the above expression, it is easy to recover the symmetry $\chi_+(\betab-\alphab) = \chi_+(\alphab)$ of \Cref{theo_AGMT} for $\alphab\in\rr^\ell$. Set $D(\xi)=S^*(\xi)\,\e^{-K(\betab- \alphab,\,\xi)}\, S(\xi) \e^{-K(\alphab,\,\xi)} $. Since $\chi_+(\alphab)\in \rr$, then $\det (\id+D(\xi))=\det (\id+D(\xi)^*)$. Using $S^*=JSJ$, $J^2=1$ and $\e^{-K(\alphab,\,\xi)} J=J\e^{-K(\alphab,\,\xi)}$
the symmetry follows immediately.

Last, the translation symmetry $\chi_+(\alphab+\theta{\bm 1}) = \chi_+(\alphab)$ follows from 
$\e^{-K(\alphab+\theta {\bm 1},\,\xi)}=\e^{\theta \epsilon(\xi)\id}\e^{-K(\alphab,\,\xi)}$.

\subsection{Open XY chain}

We first describe the general XY chain over a finite discrete interval $[a,b]\subset \zz$ (in all of this section, the notation $[\cdot,\cdot]$ refers to discrete intervals). The Hilbert space is $\cH_{[a,b]}=\big(\cc^2\big)^{\otimes [a,b]}$, with Hamiltonian
\[
H_{[a,b]}=-\frac{J}{4}\sum_{a \leq x<b}(\sigma_x^{(1)}\sigma_{x+1}^{(1)}
+\sigma_x^{(2)}\sigma_{x+1}^{(2)})-\frac\lambda2 \sum_{a\leq x\leq b} \sigma_x^{(3)}.
\]
where $\sigma_x^{(i)}$ acts on the $x$-th copy of $\cc^2$ as the Pauli matrix $\sigma^{(i)}$. This describes a spin system, where spins are localized at sites $x\in [a,b]$, with nearest neighbour coupling and magnetic field in the $(3)$-direction. The constants $J$ and $\lambda$ represent the strengths of, respectively, the nearest neighbor coupling and the magnetic field.

Now fix $M\in \nn$. For any integer $L>M$, consider the above model with $[a,b]= [-L,+L]$. 
We view the sites belonging to $[-M,+M]$ as the small system, and sites belonging to $[M+1,L]$ (respectively to $[-L,-M-1]$) as the confined right (respectively left) reservoir. With a slight abuse of notation, we denote
\[ H_\cS = H_{[-M,+M]} \qquad H_1\iL = H_{[-L,-M-1]} \qquad H_2\iL=H_{[M+1,L]},\qquad H_0\iL=H_\cS +H_1\iL +H_2\iL,\]
and 
\[
V_1=-\frac{J}{4} (\sigma_{-M-1}^{(1)}\sigma_{-M}^{(1)} +\sigma_{-M-1}^{(2)}\sigma_{-M}^{(2)})
\qquad
V_2=-\frac{J}{4} (\sigma_{M}^{(1)}\sigma_{M+1}^{(1)} +\sigma_{M}^{(2)}\sigma_{M+1}^{(2)}),
\]
so that $H\iL= H_0\iL + V_1+ V_2$. We assume that the initial state of the system is of the form
\[\rho\iL = \frac{\e^{-\beta_1 H_1\iL}}{Z_1\iL{}} \otimes \rho_\cS \otimes \frac{\e^{-\beta_2 H_2\iL}}{Z_2\iL}.\]

Generating functions for this model can be computed by identifying unitarily the XY chains with an electronic black box model using the well-known Jordan-Wigner transformation (see \cite{JOPP} for details) up to an irrelevant additive constant. In the Jordan-Wigner representation, the decoupled system is a free Fermi gas with one-particle Hilbert space
\[\ell^2([-L,+L])=\ell^2([-L,-M-1])\oplus\ell^2([-M,+M])\oplus\ell^2([M+1,L])\]
and the one-particle uncoupled and coupled Hamiltonians are
\begin{gather*}
h_0\iL=h_{[-L,-M-1]}\oplus h_{\Lambda_{[-M,+M]}}\oplus h_{[M+1,L]},\qquad 
h\iL=h_{[-L,+L]}=h_0\iL+v_1\iL+v_2\iL,
\end{gather*}
where the coupling terms
\[
v_1\iL=\frac J2\left(|\delta_{-M-1}\rangle\langle\delta_{-M}|+|\delta_{-M}\rangle\langle\delta_{-M-1}|\right),
\qquad
v_2\iL=\frac J2\left(|\delta_M\rangle\langle\delta_{M+1}|+|\delta_{M+1}\rangle\langle\delta_M|\right),
\]
are finite-rank operators. It then follows from Section \ref{sec-EBB} that our general assumptions \TL, \Blbt,
\LT\ and \LR\ are satisfied for any $\alpha_0,\theta_0$. Formulas for the generating functions are special cases of the ones obtained in the previous subsection. In particular, for $\chi_{+}(\alphab )$ starting from (\ref{eq_ebb model2}) and using the explicit form of the scattering matrix
\[s(\xi)=\e^{- 2\i\, \mathrm{sign}(J)M\xi}\begin{pmatrix}0&1\\1&0\end{pmatrix},\]
one has, denoting $\alphab=(\alpha_1, \alpha_2)$ and $\betab=(\beta_1, \beta_2)$,
 \begin{align*}
 \label{eq-spinXY}
\chi_{+}(\alphab )
&=\frac{1}{2\pi}\int_{0}^{2}
\log\frac{ \cosh\frac u2\,{(\beta_1-\alpha_1 +\alpha_2)}\, \cosh\frac u2{(\alpha_1 +\beta_2-\alpha_2)} } 
{\cosh(\frac12{\beta_1 u})\cosh(\frac12{\beta_2 u})}\,\d u.
\end{align*}
In addition, the Jordan-Wigner transformation shows that the XY-chain model is time-reversal invariant. Therefore \Cref{theo_refinement1stlaw2,theo_firstlaw_avg,theo_cvgTCL,theo_SecondLaw3,theo_onsager} hold for the XY-chain.
Note that symmetries $\chi_+(\alphab+\theta{\bm 1}) = \chi_+(\alphab)$ and $\chi_+(\betab -\alphab) = \chi_+(\alphab)$ are apparent in the above expression.

\section{Proofs of the bounds for confined systems}\label{sec_proofs_bounds}
In this section we gather the technical proofs for the bounds in \Cref{subsec_bounds}.

\subsection{Trace and norm inequalities}
In this section we give two relevant general inequalities which will be used to prove the bounds in \Cref{subsec_bounds}.
\begin{lemma}
Let $A$ and $X$ be two operators on a Hilbert space, with $A$ bounded and $X$ trace-class. We have 
\begin{equation}\label{eq_trAB}
|\tr (AX)| \leq \|A\|\, \tr |X|.
\end{equation}
\end{lemma}

\proof
Let $X=U|X|$ be the polar decomposition of $X$ and $(\psi_j)_j$ be an orthonormal basis of eigenvectors for $|X|$, with corresponding eigenvalues $(|x_j|)_j$. We have 
\[|\tr (AX)| = \sum_j |\langle \psi_j , A \, U |X|\, \psi_j \rangle| \leq \sup_j |\langle \psi_j , AU \, \psi_j \rangle |\, \sum_j |x_j| \]
which implies $\tr (AX)\leq \|A\|\, \tr |X|$. \qed

\begin{lemma}
For any two bounded $A$ and $B$ one has
\begin{equation}
\label{eq_eABeA}
\|\e^{A+B}\e^{-A}\|\leq\exp \sup_{s\in[0,1]}\|\e^{sA}B\e^{-sA}\|.
\end{equation}
\end{lemma}
 
 \proof
One has the obvious relations
 \[ \frac{\d}{\d s}\|\e^{s(A+B)}\,\e^{-s A}\|\leq \|\frac{\d}{\d s}\,\e^{s(A+B)}\,\e^{-s A}\|=\|\e^{s(A+B) }B\,\e^{-s A}\|\leq \|\e^{s(A+B)}\,\e^{-s A}\|\, \|\e^{+sA}B\,\e^{-s A}\|,
 \]
 so that \[\log \|\e^{(A+B)}\,\e^{-A}\|\leq \int_0^1 \|\e^{+sA}B\,\e^{-s A}\|\, \d s \]
 and the result follows.
\qed
 
\subsection{Proofs for Section \ref{subsec_bounds}}\label{subsec_proof_subsec_bounds}
We first prove \Cref{prop_energycorr2}; note that the upper bound and inequality \eqref{eq_CauchySchwarz} imply the lower bound.
Since $\rhot$ and $\alphab.\bm E$ commute, it follows from relation \eqref{eq_defcorr2} that 
\begin{equation} \label{eq_chitalphaGamma}
\chi_t(\alphab)=\tr\big(\rhot\,\e^{+\i t (H_0+V_{\alphab})}\,\e^{-\i t (H_0+V_{-\alphab})}\big).
\end{equation}
Remark that $\e^{+\i t (H_0+V_{\alphab})}$ and $\e^{-\i t (H_0+V_{-\alphab})}$ are mutually adjoint. Using \eqref{eq_trAB} we have
\[
\chi_t(\alphab) \leq \|\e^{+\i t (H_0+V_{\alphab})}\,\e^{-\i t (H_0+V_{-\alphab})}\| \leq \|\e^{+\i t (H_0+V_{\alphab})}\,\e^{-\i t H_0}\|^2.
\]
Using \eqref{eq_eABeA} with $A=\i t H_0$ and $B=\i t V_\alphab$ we have
\begin{align*}
|\chi_t(\alphab)|
&\leq \exp \big( 2 |t| \sup_{s\in[0,1]} \|\e^{+\i s t H_0} V_\alphab \,\e^{-\i s t H_0}\|\big)\\
&= \exp \big(2 |t| \|V_\alphab \|\big).
\end{align*} This concludes the proof of \Cref{prop_energycorr2}.
\smallskip

We now turn to the proof of \Cref{cut}. Recall that $V_\alphab$ was defined in \eqref{eq_defValpha}. We first prove that the similar upper bound 
\begin{equation} \label{eq_preuveAGMT}
  \chi_t({\alphab} + \theta{\bm 1}) \leq \chi_t({\bm \alpha})\, \e^{+|\theta|\, T_\betab({\alphab},\theta)},
\end{equation}
where
\[T_\betab(\alphab,\theta)=\sup_{ s\in[0,1]} \|V_{\alphab+s\theta{\bm 1}}\|+\sup_{ s\in[0,1]} \| V_{\betab-\alphab-s\theta{\bm 1}}\|,\]
is valid for any $\alphab$ (not necessarily satisfying $\alphab.\bm 1=0$) and $\theta$. We have $\rho=\rhot=Z^{-1}\,\e^{-\bm{\beta}.\bm E}$. Let
\[\Gamma(\alphab,\theta)=\e^{\frac12 \theta(H_0+V_{\alphab})}\e^{-\frac12 \theta H_0}=\e^{\frac12\alphab.\bm E}\e^{\frac12 \theta H}\e^{-\frac12\theta H_0}\e^{-\frac12 \alphab.\bm E}\]
and
\[D(\alphab,\theta)=\Gamma(\alphab,\theta)\Gamma^*(\alphab,\theta).\]
Starting from \eqref{eq_defcorr2},
\[\chi_t(\alphab+\theta\bm 1)=Z^{-1}\tr\big(D(\alphab,\theta)\e^{-\frac12\alphab.\bm E}\e^{-\i t H}\e^{-\frac12(\betab-\alphab).\bm E}D(\betab-\alphab,-\theta)\e^{-\frac12(\betab-\alphab).\bm E}\e^{+\i t H}\e^{-\frac12 \alphab.\bm E}\big).\]
Applying twice inequality \eqref{eq_trAB} we have:
\begin{align*}
\chi_t(\alphab+\theta {\bm 1})&\leq Z^{-1} \|D(\alphab,\theta)\|\tr\big(\e^{-\frac12\alphab.\bm E}\e^{-\i t H}\e^{-\frac12(\betab-\alphab).\bm E}D(\betab-\alphab,-\theta)\e^{-\frac12(\betab-\alphab).\bm E}\e^{+\i t H}\e^{-\frac12 \alphab.\bm E}\big)\\[2mm]
&= Z^{-1} \|D(\alphab,\theta)\|\tr\big(D(\betab-\alphab,-\theta)\e^{-\frac12(\betab-\alphab).\bm E}\e^{+\i t H}\e^{-\alphab.\bm E}\e^{-\i t H}\e^{-\frac12(\betab-\alphab).\bm E}\big)\\[2mm]
&\leq Z^{-1} \|D(\alphab,\theta))\|\|D(\betab-\alphab,-\theta)\|\tr\big(\e^{-\frac12(\betab-\alphab).\bm E}\e^{+\i t H}\e^{-\alphab.\bm E}\e^{-\i t H}\e^{-\frac12(\betab-\alphab).\bm E}\big)\\[2mm]
&= \|D(\alphab,\theta)\|\|D(\betab-\alphab,-\theta)\|\,\chi_t(\alphab).
\end{align*}
From \eqref{eq_eABeA}, for any $\alphab\in\rr^\ell$ and $\theta\in\rr$,
\begin{align*}
\|D(\alphab,\theta)\|&=\|\e^{\frac12 \theta(H_0+V_{\alphab})}\e^{-\frac12 \theta H_0}\|^2\leq \exp(|\theta|\sup_{s\in[0,1]}\|V_{\alphab+s\theta\bm 1}\|),\\
\|D(\betab-\alphab,-\theta)\|&=\|\e^{-\frac12 \theta(H_0+V_{\betab-\alphab})}\e^{\frac12 \theta H_0}\|^2\leq \exp(|\theta|\sup_{s\in[0,1]}\|V_{\betab-\alphab-s\theta\bm 1}\|).
\end{align*}
The desired inequality therefore holds. In addition, the inequality $T_\betab(\alphab,\theta)\leq S_\betab(\norme{\alphab},|\theta|)$ holds when $\alphab.\bm 1=0$. This yields the upper bound in \eqref{eq_preuveAGMT}. The obvious symmetry $T_\betab(\alphab,\theta)=T_\betab(\alphab+\theta \bm 1,-\theta)$ gives the lower bound.

\appendix

\section{The role of the small system} \label{subsec_smallsystem}

We consider a system which we view as consisting of a fixed system $\sys$ described by $(\cH_\sys,\rho_\sys, H_\sys)$, which is coupled to reservoirs $\cR_j$, $j=1,\ldots,\ell$, each of which is decribed by $(\cH_j^{(L)}, \rho_j\iL, H_j\iL)_{L\in\nn}$. We assume that $\cH_\sys$ is of finite dimension, and that $\cH_\sys,\rho_\sys, H_\sys$ do not depend on $L$. Then the $L$-th confined full system is described by the Hilbert space $\cH_f^{(L)}=\cH_\sys\otimes \cH^{(L)}$, with the free Hamiltonian $H_\sys\otimes \id+ \id\otimes H_0$ and initial state $\rho_f^{(L)}=\rho_\sys\otimes \rho^{(L)}$. The coupling is described by an observable $V^{(L)}$ on $\cH_f^{(L)}$ so that the full Hamiltonian is given by $H_f^{(L)}= H_\sys+ H_0^{(L)}+V^{(L)}$.

One way to look at this system is to consider the small system $\sys$ as an $\ell+1$-th reservoir and measure the energies both in the reservoirs $\cR_j$ and in the small system $\sys$. We call this the ``full'' description and label corresponding objects with an $f$. It amounts to consider the multi-reservoir system $(\cH_f, \rho_f\iL,\bm E_f\iL , V\iL)_{L}$ with $\bm E_f\iL=(H_\sys, H_1^{(L)},\ldots,H_\ell^{(L)})$. Setting $\alphab_f=(\alpha_\sys, \bm \alpha)$, we denote by $\chi_{f,t}\iL(\alphab_f)$ the associated generating function on $\rr\times \rr^\ell$. Another way to look at the system is to view $\sys$ as part of the interaction and therefore to consider only the $\ell$ reservoirs. We call this the ``reduced'' description and label corresponding objects with an $r$. It amounts to consider the multi-reservoir system $(\cH_f, \rho_f\iL, \bm E, H_\sys+V^{(L)})_{L}$. We have:
\begin{align*}
\chi_{f,t}\iL(\alphab_f)&=\tr \big(\e^{-\i t H\iL}{\tilde\rho_\sys\otimes \tilde\rho\iL} \,\e^{+\alpha_\sys H_\sys+{\alphab}. {\bm E^{(L)}}}\e^{+\i t H\iL}\e^{-\alpha_\sys H_\sys-{\alphab}. {\bm E^{(L)}}} \big),\\
\chi\iL_{r,t}(\bm \alpha)&= \tr \big(\e^{-\i t H\iL}{\rho_\sys \otimes \tilde\rho\iL} \,\e^{+{\alphab}. {\bm E^{(L)}}}\e^{+\i t H\iL}\e^{-{\alphab}. {\bm E^{(L)}}} \big)
\end{align*}
and obviously, 
\begin{equation} \label{eq_inegchirchif}
  \e^{-2|\alpha_\sys| \norme{H_\sys}} \chi_{r,t}(\alphab)\leq \chi_{f,t}\iL(\alpha_\sys,\alphab)\leq \e^{2\module{\alpha_\sys} \norme{H_\sys}} \chi_{r,t}(\alphab),
\end{equation}
so that, even though the assumption \TL\ in the reduced and the full picture are not equivalent (although the latter implies the former), as soon as they are assumed to hold then $\overline\chi_{r,+}(\alphab)=\overline\chi_{f,+}(\alpha_\sys,\alphab)$ for all $\alpha_\sys$, and both versions of \LT\ are equivalent and define the same functional $\chi_+$, which only depends on $\alphab$. Similarly, the fact that $H_\sys$ commutes with the $H_j$, $j=1,\ldots,\ell$ implies that assumptions \Blt\ and \Blbt\ are equivalent for the full and the reduced pictures. Therefore, the choice of the full or reduced description is irrelevant to the conclusions of all our large-time results results, from \Cref{theo_AGMT} to \Cref{theo_onsager}.

\section{Proofs for Section \ref{sec-models}} 
\label{sec:proofs_for_section_yyy}
For the spin system, of \Cref{subsec:spinsys}, since the considered interaction has finite range, a direct application of e.g.\ Theorem 6.2.4 in \cite{BR2} shows the existence of strongly continuous one-parameter groups $(\tau^t)_t$, $(\tau_{\bm{\alpha}}^t)_t$ of *-automorphisms on the norm closure ${\mathcal O}=\overline{\bigcup_{X\subset{\mathcal G}}\bigotimes_{x\in X}M_k(\cc)}$, such that for $\alphab\in\cc^\ell$ with $\norme\alphab$ small enough,
\begin{equation} \label{eq_QSScvg}
 \lim_{L\to\infty} \|\e^{+\i t H\iL}A \,\e^{-\i t H\iL}- \tau^t(A)\|=0\qquad \lim_{L\to\infty} \|\e^{+\i t \alphab.\bm E\iL}A \,\e^{-\i t \alphab.\bm E\iL}- \tau_{\alphab}^t(A)\|=0 
\end{equation}
uniformly for $t$ in a compact set of $\rr_+$ and $A\in\mathcal O$. We will denote by $\tau^t_0$ the map $\tau^t_{\alphab}$ for ${\alphab}={\bm 1}$, which is the limit of the evolution associated with the free Hamiltonian $H_0\iL$. 

\paragraph{Proof of \Cref{thm-TDL spins}}
An immediate adaptation of the proof of Theorem 6.2.4 in \cite{BR2} shows that there exists two constants $\gamma_0>0$ and $C>0$ such that for any $x\in \cup_{j=1}^\ell \partial\mathcal G_j$, any $\alphab$ with $\norme\alphab<\gamma_0$, one has
\begin{equation} \label{lemma_BR2QSS}
 \limsup_{L}\|\e^{+\frac12 \alphab.\bm E\iL }S_x\,\e^{-\frac12 \alphab.\bm E\iL }\| \leq C \,\|A\|.
\end{equation}
Expanding $V\iL$ and applying \Cref{lemma_BR2QSS} implies \Blt\ provided $\theta_0<\gamma_0$.

We now prove \TL. Let $\Gamma_\alphab\iL(t)=\e^{+\i t (H_0\iL+V\iL_\alphab)}\e^{-\i t H_0\iL}$, with $V\iL_\alphab=\e^{+\frac12\alphab.\bm E\iL} V\iL \e^{-\frac12\alphab.\bm E\iL}$ as before. Fix $\alphab\in\i \rr^\ell$; from relation \eqref{eq_chitalphaGamma} we have 
  \begin{equation} \label{eq_chitalphagammaalpha}
  \chi_t\iL(\alphab)= \tr \big(\rho\iL\, \Gamma_\alphab\iL(t)\,\Gamma_\alphab\iL(t){}^*\big).
  \end{equation}
  Let $V$ and $V_\alphab$ be the (norm) limits as $L\to\infty$ of $V\iL$ and $V\iL_\alphab$ respectively. Their existence is assured by our assumption on $x\mapsto J(x)$, and $\alphab \in \i \rr^\ell$. Let $\Gamma_\alphab(t)$ be as the solution of $\frac\partial {\partial t}\, \Gamma_\alphab(t)=\i\Gamma_\alphab(t)\,\tau_0^t(V_\alphab)$, with initial condition $\Gamma_\alphab(0)={\id}$ (this $\Gamma_\alphab(t)$ can be explicitly constructed in terms of a Dyson expansion).
   By differentiating $\Gamma_\alphab^{(L)}(t)\,\Gamma_\alphab^*(t)$, and using the unitarity of $\Gamma_\alphab^{(L)}(t)$ and $\Gamma_\alphab^*(t)$, we obtain
  \[\|\Gamma_\alphab^{(L)}(t)\Gamma_\alphab^*(t)- \id\| \leq \int_0^t\| \e^{+\i s H_0\iL}V\iL_\alphab \,\e^{-\i s H_0\iL}-\tau^s_0(V_\alphab) \|\, \d s.\]
  Since by \Blt\ and \eqref{eq_QSScvg}, $\| \e^{+\i s H_0\iL}V\iL_\alphab \,\e^{-\i s H_0\iL}-\tau^s_0(V_\alphab)\|$ is uniformly bounded and converges to~ $0$ as $L\to\infty$ for any $t$, we have $\lim_{L\to\infty }\|\Gamma_\alphab^{(L)}(t)-\Gamma_\alphab(t)\|=0$.   Relation \eqref{eq_chitalphagammaalpha} and the assumption that the states converge then imply
  \[\lim_{L\to\infty} \chi_t^{(L)}(\alphab)= \rho\big(\Gamma_\alphab(t)\Gamma_{-\alphab}^*(t)\big)\]
  uniformly for $t$ in any compact set, and $\alphab\in\i \rr^\ell$. Continuity of $\chi_t(\alphab)$ in $\alphab=0$ along $\i \rr^\ell$ is a consequence of the uniform in $\alphab\in\i\rr^\ell$ norm convergence of the Dyson expansion of $\Gamma_\alphab(t)$. This proves \TL. \hfill\qed

\paragraph{Proof of Proposition \ref{prop_TLBspinfermion}}

We have
\begin{equation*}
\chi_{t} \iL(\alpha_\cS,\alphab)= \frac12 \tr \Big( \e^{-\sum_j \beta_j H_j\iL} \, \e^{+\i t H\iL_{\alpha_\sys,\alphab}}\, \e^{-\i t H\iL} \Big)/ \Big( \prod_{j=1}^\ell Z\iL_j \Big)
\end{equation*}
with
\begin{equation*}
H\iL_{\alpha_\sys,\alphab} =H_\cS + \sum_j H_j + \lambda \sum_j \begin{pmatrix}0 & \e^{+2\alpha_\cS} \\ \e^{-2\alpha_\cS} & 0 \end{pmatrix} \otimes \frac1{\sqrt 2}\big( a^*_j (\e^{+\alpha_j h_j\iL}v_j\iL)+a_j (\e^{-\alpha_j h_j\iL}v_j\iL)\big) 
\end{equation*}
A Dyson expansion of $ \e^{+\i t H\iL_{\alpha_\sys,\alphab}}\, \e^{-\i t H\iL}$ shows that $\chi_{t} \iL(\alpha_\cS,\alphab)$ converges for any $(\alpha_\cS,\alphab)$ in $\i\rr^{\ell+1}$ as $L\to\infty$ (see \cite[Appendix B]{BPR} for some related techniques of thermodynamic limit on Fock spaces). 

\paragraph{Proof of \Cref{prop_SFB}} 
\label{par:proof_of_prop_SFB}

Using $\|a^\#(v)\|\leq \|v\|$ and the positivity of each $h_j\iL$, one has for $\alphab$ in $\cB(\alpha_0)$ and $\theta$ in~$[-\theta_0,+\theta_0]$
\[\|\e^{+\frac12(\alphab+\theta \bm 1). \bm E\iL} V\iL \e^{-\frac12(\alphab+\theta \bm 1). \bm E\iL}\|\leq |\lambda|\e^{(\alpha_0+\theta_0)}\sum_{j=1}^\ell \big(\|\e^{+\frac12(\alpha_0+\theta_0) h_j\iL} v_j\iL\|+\|v_j\iL\|\big).\]
Assumption \SFUV\ therefore implies \Blt\ whenever $\alpha_0\leq\theta_0\leq \gamma_0$. The proof regarding \Blbt\ is similar.




\section{Analytic approximation of the interaction} \label{sec_analyticapprox}
 
As discussed in \Cref{sec-setup-general}, starting from any multi-reservoir system $(\cH^{(L)}, \rho^{(L)},\bm E^{(L)},V^{(L)})_{L\in\nn}$ we can always find a sequence $(\tilde V^{(L)})_{L\in\nn})$ approximating $(V^{(L)})_{L\in\nn}$ uniformly in $L$, and such that \Blbt\ is true with any $\alpha_0$ and $\theta_0$ in $(0,+\infty)$, provided that $t\mapsto\e^{+\i\alphab.\bm E^{(L)}}V^{(L)}\e^{-i\alphab.\bm E^{(L)}}$ fulfills some uniform (in $L$) continuity condition at~$t=0$. More precisely, we have:
 \begin{proposition}
 \label{prop-analytic-approx}
 Assume that $\sup_{L} \|V^{(L)}\|<\infty$ and 
 \begin{equation}
 \lim_{\|\alphab\|\to 0}\sup_{L\in\nn} \|\e^{+\i\alphab.\bm E^{(L)}}V^{(L)}\e^{-\i\alphab.\bm E^{(L)}}-V^{(L)}\|=0.
 \end{equation}
Then for any $\varepsilon>0$ there exists a sequence of $(\tilde V_\varepsilon^{(L)})_{L\in\nn}$, with $\tilde V_\varepsilon^{(L)}=\tilde V_\varepsilon^{(L)}{}^*\in \mathcal B(\mathcal H^{(L)})$, such that
 \begin{equation}
 \sup_{L} \|\tilde V_\varepsilon^{(L)}-V^{(L)}\|<\varepsilon
 \end{equation}
and the multi-reservoir system $(\cH^{(L)}, \rho^{(L)},\bm E^{(L)},\tilde V_\varepsilon^{(L)})_{L\in\nn}$ satisfies assumption \Blbt\  for any $\betab \in \rr^\ell$, for any $\alpha_0$ and $\theta_0$ in $(0,+\infty)$,
 \end{proposition}

\proof
 We define (denoting $\d\bm\sigma=\d\sigma_1\ldots \d \sigma_\ell$)
 \begin{equation*}
 \tilde V^{(L)}_{N}=\sqrt{\tfrac{N}{\pi}}\int_{\rr^\ell} \e^{+\frac12 \i\bm\sigma.\bm E^{(L)}} V^{(L)} \e^{-\frac12\i\bm\sigma.\bm E^{(L)}}\,\e^{-N \sum_{j=1}^\ell \sigma_j^2} \,\d \bm\sigma.
 \end{equation*}
 This is the analytic element defined in \cite[Proposition 2.5.22]{BR1} .
 As a direct consequence of this definition, for any $L$ and $N$ in $\nn$, we have $\|\tilde V_N^{(L)}\|\leq \|V^{(L)}\|$. In addition, it is easy to prove that for any $\varepsilon>0$, there exists $N_\varepsilon>0$ such that $\sup_{L}\|\tilde V_{N_\varepsilon}^{(L)}-V^{(L)}\|<\varepsilon.$
We now drop the subscript $N_\varepsilon$ for $\tilde V_{N_\varepsilon}^{(L)}$ and we show that for any $\alphab$ in $(0,+\infty)^\ell$, one has $\sup_{L}\norme{\tilde V^{(L)}_\alphab}<\infty$ where $\tilde V^{(L)}_\alphab$ is $V^{(L)}_\alphab$ with $V^{(L)}$ replaced by $\tilde V^{(L)}$.

 For $\alphab \in \i\rr^\ell$ a simple change of variable gives
 \begin{equation*}
 \tilde V\iL_\alphab=\sqrt{\tfrac{N}{\pi}} \int_{\rr^\ell} \e^{+\frac12 \i \bm\sigma.\bm E^{(L)}}V^{(L)}\e^{-\frac12 \i \bm\sigma.\bm E^{(L)}} \, \e^{-N(\bm{\sigma}+\i\alphab).(\bm{\sigma}+\i\alphab)}\, \d \bm \sigma.
 \end{equation*}
 
 For a general $\alphab\in \cc^\ell$, the integrand on the right end side and its derivative in $\alphab$ can be 
 norm bounded by 
\[\sup_L \|V\iL\|\, \big(1+2N(\|{\bm \sigma}\|+\|\alphab\|)\big)e^{N\sum_{j=1}^\ell(\Re \alpha_j)^2}\e^{-N \sum_{j=1}^\ell (\sigma -\Im \alpha_j)^2},\]
so the integral is well defined for any $\alphab\in \cc^\ell$ and the map $\alphab\in\cc^\ell \mapsto I{\iL}(\alphab)\in \cB(\cH)$ it defines is entire analytic. 
Since for any $L$ finite $\alphab\mapsto \tilde V\iL_\alphab$ is entire analytic, $\tilde V\iL_\alphab=I\iL(\alphab)$ for any $\alphab\in \cc^\ell$. Since $I\iL$ is uniformly bounded in $L$ and $\alphab$ on any compact subset $K \subset \cc^\ell$, so is $\alphab\mapsto \tilde V\iL_\alphab$ and \Blbt\ is satisfied for any $\betab, \alpha_0, \theta_0$.
\qed
 
\begin{remark}
 We would like to insist that this proposition does not imply that \Blbt\ is fulfilled for $V^{(L)}$, even if $V^{(L)}$ is the limit of an analytic approximation such as $\tilde V^{(L)}$: in general we do not have
 \[\lim_{\varepsilon \to 0} \sup_{L} \|\tilde V^{(L)}_{\varepsilon, \alphab} -V^{(L)}_\alphab\|=0\]
 since the factor $\e^{N\sum_{j=1}^\ell(\Re(\alpha_j))^2}$ blows up as $N$ grows to infinity.
\end{remark}


\newcommand{\etalchar}[1]{$^{#1}$}
\providecommand{\MR}[1]{} \newcommand{\noop}[1]{}
\providecommand{\bysame}{\leavevmode\hbox to3em{\hrulefill}\thinspace}
\providecommand{\MR}{\relax\ifhmode\unskip\space\fi MR }
\providecommand{\MRhref}[2]{%
	\href{http://www.ams.org/mathscinet-getitem?mr=#1}{#2}
}
\providecommand{\href}[2]{#2}

\end{document}